%% file: paper.tex
\documentclass[journal]{IEEEtran}
\usepackage{xcolor}

\usepackage{colortbl}
\usepackage{diagbox}
\usepackage{makecell}
\usepackage{booktabs}
\usepackage{multirow}

\usepackage{amsmath,amsfonts}
\usepackage{bm}
\usepackage{tikz}
\usepackage{cases}
\usepackage{amsthm}
\usepackage{upgreek}
\usepackage{amssymb}

\usepackage[linesnumbered,algoruled,lined]{algorithm2e}

\usepackage{enumerate}
\usepackage{verbatim}
\usepackage{textcomp}
\usepackage{comment}
\usepackage{subcaption}

\usepackage{enumitem}

\newtheorem{theorem}{Theorem}
\newtheorem{remark}{Remark}
\newtheorem{assumption}{Assumption}
\newtheorem{definition}{Definition}

\newcommand{\defeq}{\stackrel{\mathrm{def}}{=}}

\begin{document}

\title{PPVF: An Efficient Privacy-Preserving Online Video Fetching Framework with Correlated Differential Privacy}

\author{Xianzhi Zhang,
        Yipeng Zhou,
        Di Wu,
        Quan Z. Sheng, 
        Miao Hu, 
        and Linchang Xiao
\thanks{
\IEEEcompsocthanksitem Xianzhi Zhang, Di Wu, Miao Hu, Linchang Xiao are with the School of Computer Science and Engineering, Sun Yat-sen University, Guangzhou, 510006, China, and the Key Laboratory of Machine Intelligence and Advanced Computing (Sun Yat-sen University), Ministry of Education, China. E-mail: \{zhangxzh9, xiaolch3\}@mail2.sysu.edu.cn; \{wudi27, humiao5\}@mail.sysu.edu.cn. \emph{(Di Wu is the corresponding author.)} 
\IEEEcompsocthanksitem Yipeng Zhou, and Quan Z. Sheng are with the School of Computing, Macquarie University, NSW 2109, Australia. E-mail: \{yipeng.zhou, michael.sheng\}@mq.edu.au. 
}
} 


\maketitle

\begin{abstract}
 Online video streaming has evolved into an integral component of the contemporary Internet landscape. Yet, the disclosure of user requests presents formidable privacy challenges. As users stream their preferred online videos, their requests are automatically seized by video content providers, potentially leaking users' privacy. 
 Unfortunately, current protection methods are not well-suited to preserving user request privacy from content providers while maintaining high-quality online video services. 
To tackle this challenge, we introduce a novel Privacy-Preserving Video Fetching (PPVF) framework, which
utilizes trusted edge devices to pre-fetch and cache videos, ensuring the privacy of users' requests while optimizing the efficiency of edge caching.
More specifically, 
we design PPVF with three core components: (1) \textit{Online privacy budget scheduler}, which employs a theoretically guaranteed online algorithm to select non-requested videos as candidates with assigned privacy budgets. Alternative videos are chosen by an online algorithm that is theoretically guaranteed to consider both video utilities and available privacy budgets. 
(2) \textit{Noisy video request generator}, which generates redundant video requests (in addition to original ones) 
utilizing correlated differential privacy to obfuscate request privacy. 
(3) \textit{Online video utility predictor}, which leverages federated learning to collaboratively evaluate video utility in an online fashion, aiding in video selection in (1) and noise generation in (2). 
Finally, we conduct extensive experiments using real-world video request traces from Tencent Video. The results demonstrate that PPVF effectively safeguards user request privacy while upholding high 
video caching performance.
\end{abstract}

\input{body}

\bibliographystyle{IEEEtran}
\bibliography{IEEEabrv,reference}


\input{bio}

\input{appendix}

\end{document}

%% file: body.tex
\section{Introduction}
\IEEEPARstart{O}{nline} video streaming 
has become 
an indispensable service in our daily lives, serving billions of Internet users by streaming diverse videos, including movies, news, and TV episodes. In 
2023, YouTube alone provided over 5 billion online videos daily to more than 122 million Internet users, with a total daily playback time exceeding 1 billion hours~\cite{GlobalMediaInsight2022}. To serve such a huge number of users, it is common to leverage edge devices (EDs) to pre-fetch video content for users. Such EDs, including user devices~\cite{Nisha2022, Amini2011, Wang2019}, mobile vehicles~\cite{Qian2020, Hu2018, Cui2020}, access points (APs)~\cite{Wang2019, Zhang2022b}, and edge nodes of the content delivery network (CDN)~\cite{Qiao2022}, can significantly reduce communications between users and video content providers (CPs). Meanwhile, serving users with content cached on EDs can achieve a high quality of service (QoS) and a low playback latency for video streaming.

However, with the proliferation of the online video market, the  \emph{request privacy leakage} concern has risen~\cite{Cui2020, Sivaraman2021, Tong2022}.
When users request videos from online CPs, their request traces are automatically recorded by CPs. Analyzing these traces can potentially reveal sensitive privacy information such as gender~\cite{Qiao2022}, age~\cite{Zhou2019}, location~\cite{Nisha2022, Hu2018}, and hobbies~\cite{Cai2023, Yang2016}.
The leakage of request privacy poses significant risks to users with the spread of spam and scams~\cite{Xiao2018}. Consequently, it is urgent and essential to develop effective video request strategies with preserved user privacy~\cite{Ni2021}. 

Various privacy-preserving methodologies, such as encryption, federated learning (FL) and differential privacy (DP), have emerged, 
but applying them to preserve video request privacy
is non-trivial, which can be explained as follows. 
Encryption-based methods, e.g., Hypertext Transfer Protocol Secure (HTTPS)~\cite{Araldo2018}, can only shield against privacy threats from external attackers when delivering videos.
FL is a generic framework to preserve privacy by only exposing parameters when training machine learning models for edge devices~\cite{Liu2024, Nguyen2024}. 
Yet, they fail to conceal request traces from CPs because user requests must remain visible to CPs for 
video streaming services
to function properly. 
DP is a widely utilized approach to safeguarding user privacy in machine learning systems~\cite{Zhou2019, Liu2023, Hu2024}. 
However, straightly distorting user requests with DP noises can severely impair video streaming efficiency by requesting many videos out of users' interest.

To mitigate video request privacy leakage, we propose a novel Privacy-Preserving Video Fetching (PPVF) framework by synthetically utilizing video pre-fetching, FL and correlated differential privacy (CDP) to supplement each other 
and 
overcome the deficiency of each single methodology. 
For deployment in practice, our framework can be implemented on trusted EDs, \emph{e.g.}, user devices~\cite{Nisha2022, Amini2011, Wang2019}, vehicles~\cite{Qian2020, Hu2018, Cui2020}, access points (APs)~\cite{Wang2019, Zhang2022b}, owned or trusted by users.
By utilizing the PPVF framework, trusted EDs achieve a harmonious blend of efficient video delivery and safeguard the privacy of viewing records.
In summary, our main contributions are listed below:
\begin{itemize}
    \item To the best of our knowledge, we are among the first to propose an \textit{online privacy budget scheduler} for reconciling privacy and efficiency in online video systems. Candidate videos with assigned privacy budgets are tactfully selected by a threshold-based online algorithm for further distorting the process.  
    Additionally, the performance of the allocation algorithm is theoretically guaranteed by competitive analysis.
    \item We leverage Correlated Differential Privacy (CDP) to design the \textit{noisy video request generator} for generating video requests (including redundant noisy ones) in edge video caching systems.
    By taking correlated video request patterns into account, CDP can more accurately calibrate the noise scale when distorting pre-fetching requests and hence avoid injecting excessive noises. 
    \item To predict video utility (serving as the pivotal prior knowledge for video selection and noise generation), we further construct a FL-based \textit{online video utility predictor}, which only exposes non-critical parameters to collaboratively evaluate video utility in an online and privacy-preserving mode. 
    \item We 
    conduct extensive experiments by using real-world request traces collected from Tencent Video~\cite{Zhang2022} to validate the superiority of PPVF. 
The experimental results demonstrate that PPVF is the best 
in preserving privacy without significantly compromising caching performance compared with the state-of-the-art baselines.
\end{itemize}

The remainder of the paper is organized as follows.
The PPVF system architecture, the threat model, and the 
problems formulation are 
presented 
in Section~\ref{sec: SYSTEM MODELS}. 
In Section~\ref{sec:proposed_alg}, we introduce novel algorithms for allocating the privacy budget and determining the pre-fetching strategy in edge caching systems. 
Section~\ref{sec: Evaluation Method} presents the video utility predictor obtained via federated learning and online parameter estimation methods. 
The experimental results are 
reported 
in Section~\ref{sec: Experiment} followed by a discussion of the related works in Section~\ref{Sec:Related}.
Finally, we conclude our paper in Section~\ref{sec: conclusion}.

\section{System Model and Preliminary}\label{sec: SYSTEM MODELS}

In this section, we introduce the system model of our PPVF framework and the main entities in PPVF. To facilitate readability, we have summarized notations in Table~\ref{Table:Major Notations}.

\subsection{System Architecture}
There are three types of entities in the system, which are briefly introduced as follows:
\begin{itemize}
\item 
\textbf{\textit{Content Provider (CP)}}: The CP is an online video service provider possessing a comprehensive set of $\text{I}$ videos denoted by $\mathcal{I}=\{i_1, i_2, \cdots, i_\text{I}\}$. 
However, the CP also collects users' request traces to enhance its services, e.g., recommendation~\cite{Guerraoui2017, Cai2023} and advertisement~\cite{Zhou2019, Yang2016}, which is regarded as the main risk entity in our privacy model. 
\item 
\textbf{\textit{Trusted Edge Devices (EDs)}}: In PPVF, we denote $\mathcal{E}=\{e_1,e_2,\cdots,e_\text{E}\}$ as the set of all EDs which can be trusted by users~\cite{Sen2018, Sivaraman2021,Tong2022}. 
{
Each ED $e\in\mathcal{E}$ has a storage limitation $c_e$ in our system model. 
EDs play two critical roles: i) caching videos fetched from the CP to serve users' requests with a higher quality of service (QoS), and ii) preserving video request privacy to conceal users' real preferences\footnote{Our main focus is to design a privacy framework for trusted edge devices, which may deploy Trusted Execution Environments (TEE), such as Intel SGX,  
TrustZone for Cortex-M, 
to convince users and execute instructions reliably.
}.}
\item 
\textbf{\textit{Users}}: Users are final consumers of videos. Rather than directly exposing video requests to the CP, users in PPVF only submit their requests to corresponding trusted EDs, which act as agents to fetch videos from the CP for users. 
\end{itemize}

{
\subsubsection{\textbf{Interactions Between EDs and Users}}
When a user needs to watch video content $i'\in\mathcal{I}$ at time $\tau\in[0,T)$, the user submits 
her 
request (denoted by a view content vector $\bm{v}_{e} = [v_{e,i}]^{\text{I}}$ where $v_{e,i'} = 1$ and $v_{e,i} = 0, \forall i\neq i', i\in\mathcal{I}$) to its ED $e$. 
Here, time interval $[0,T)$ is the observed window in our problem.
If the requested video is cached at ED $e$, the video can be streamed directly from the ED to the user. Otherwise, ED $e$ needs to pre-fetch some redundant videos plus the missing video $i'$ from the CP.
{
Here, we denote $k\in \mathbb{N}^+\, ,\, 1 \leq k\leq \text{K}_e$ as the index of the cache missing request that needs to be fetched from the remote CP by ED $e$, where \(\text{K}_e\) represents the maximum index number of the missing requests at ED \(e\) in $[0,T)$.}

Note that all EDs can authentically record all video requests from users to evaluate the utility for video pre-fetching and caching~\cite{Sen2018, Wang2019, Zhou2019}. For any ED $e$, let  $\mathcal{V}_{e}^k=\{\, (i, \tau) \, \mid \, i\in\mathcal{I}, \tau\in[0,t^k),
\tau\in\mathbb{R}, 1 \leq k \leq \text{K}_e\}$ denote the set of all historical viewing requests up to the time $t^k$, where $t^k$ is the time of the $k$-th cache missing request.  
Besides, we denote $\mathcal{T}_{e,i}^{t_1,t_2}$ as the timestamp set of the viewing requests for video $i$ within time window $[t_1,t_2)$ at ED $e$.

\begin{table*}[tbp]
\caption{Main notations used in the paper.}
\label{Table:Major Notations}
\rowcolors{2}{white}{gray!25} 
\begin{tabular}{p{3cm}<{\centering} p{\linewidth-3.9cm}}
\toprule
Notation & \multicolumn{1}{c}{Description}\\ 
\midrule
$\mathcal{E}$ / $\mathcal{I}$ & The space set of all edge devices (EDs) / videos. \\  
$i$ / $e$  &  The index of any video / ED. \\
$k$  & The index of request that is missed at a specific ED and needs to be fetched from the CP.\\ 
$t^{k}$ & The timestamp of the $k$-th cache missing request.\\
$\mathcal{V}_{e}$ / $\mathcal{V}_{e}^k$ & The set of all viewing requests at ED $e$ in time $[0,T)$ / $[0,t^k)$.\\
$\mathcal{T}_{e, i}$ / $\mathcal{T}_{e, i}^{t_1,t_2}$ & The timestamp set of the viewing requests of video $i$ arriving at ED $e$ in time window $[0,T)$ / $[t_1,t_2)$.\\
$\bm{x}_e^k$ / $\bm{v}_e^k$ & The pre-fetching / viewing vector for the $k$-th cache missing request at ED $e$.\\
$c_e$ / $f_e$& The caching / pre-fetching capability of ED $e$.\\
$\bm{a}_e^k$ & The privacy budget allocation vector for the $k$-th pre-fetching at ED $e$.\\
$\mathcal{A}_e^k$ & The candidate video set for generating redundant requests for the $k$-th pre-fetching at ED $e$.\\
$\xi_{e,i}$ / $\epsilon_{e,i}$ & The total privacy budget / once privacy cost of ED $e$ with respect to video $i$.\\
$\lambda_{e,i}^k$ & The utility of pre-fetching and caching video $i$ in time $t^k$ at ED $e$.\\
$\bm{\theta} = \{\bm{\beta},\bm{p},\bm{q}\}$ & The parameters of MEP model.\\
$t_{\theta}$  & The update time point of online parameter estimation based on FL-framework. \\
$\bm{\Psi}^k_e$ & The correlated degree matrix among different videos in time $t^k$ at ED $e$.\\
$\bm{\psi}_e^{k}$ / $\bm{\alpha}_e^{k}$ / $\bm{\sigma}_e^{k}$ & The historical matrices to calculate the correlated degree matrix $\bm{\Psi}^k$ in time $t^k$ at ED $e$. \\
$\Delta \lambda_{e,i}^k$ / $\Delta \lambda^k_{e,gc}$ & The correlated sensitivity for each video $i$ / global of the $k$-th pre-fetching at ED $e$.\\
\bottomrule
\end{tabular}
\vspace{-3mm}
\end{table*}

 \subsubsection{\textbf{Interactions between the CP and EDs}}
In PPVF, EDs, in lieu of end users, interact with the CP. On 
the 
one hand, the CP delivers requested videos to EDs. On the other hand, since each ED has a limited number of local viewing requests, the CP needs to assist EDs in evaluating video utility with federated learning to enhance the quality of service. 

The vector of pre-fetching requests is denoted by $\bm{x}_{e}^k= [x_{e,i}^k]^{\text{I}}$, where $x_{e,i}^k\in \{0,1\},\forall e,i,k$. To preserve privacy, ED $e$ utilizes $\bm{x}_{e}^k$ to obfuscate the original view request vector $\bm{v}_{e}^k$ for the $k$-th cache missing request, which needs to be fetched from the CP. 
Videos finally fetched by ED $e$ from the CP is conducted by $\bm{r}_{e}^k = [r_{e,i}^k]^{\text{I}}$, where $r_{e,i}^k = v_{e,i}^k |x_{e,i}^k$, representing the fetching vector sent by ED $e$ for the $k$-th cache missing request. 
Here, the symbol `$|$' represents the `OR' operator, indicating that whether ED \(e\) fetches video \(i\) depends on both \(x_{e,i}^k\) and \(v_{e,i}^k\).
Note that the generation of $\bm{v}_e^k$ is purely based on users' view interests, not affected by our strategies. 
Our study focuses on generating $\bm{x}_e^k$ for privacy protection. 


Furthermore, with the assistance of the CP, we assume that EDs can periodically update the parameters of their local model for video utility prediction without disclosing their private data. The set of time points to execute the online parameter estimation is denoted by  $t_{\theta}\in\mathcal{Q}$, where $\mathcal{Q}=\{t_{\theta}|t_{\theta}\in[0, T),t_{\theta}\in\mathbb{N}\}$. 
Note that $t_{\theta}$ 
represents the time point to update model parameters $\bm{\theta}$, not the time point for requesting videos. 
Due to the limited caching space, ED $e$ updates its cached videos by fetching videos according to predicted video utility when its cache space is full. The video utility will be further specified in Section~\ref{sec: Evaluation Method}.
}

\subsection{Threat Model}
In traditional online video systems, privacy threats related to video fetching primarily arise from the exposure of users' video-request patterns and preferences. As users interact with the CP to access the online video services, their historical video requests and pre-fetching activities will be inadvertently exposed to the CP, which can accordingly infer sensitive information, such as age~\cite{Zhou2019}, gender~\cite{Qiao2022}, locations~\cite{Nisha2022, Hu2018}, and favorites~\cite{Cai2023, Yang2016}, of users. Such threats are driven by the goal 
of enhancing 
services through caching or recommendation algorithms. CPs can exploit inferred sensitive information to gain insights into individual user preferences. Therefore, unauthorized access to user-specific information without protection poses a significant privacy threat, enabling CPs to infer personal preferences, potentially compromising users' privacy.

\subsection{Problems Formulation}


Let us first consider the global video caching problem without considering privacy leakage. 
When requesting videos missed by the edge cache from the CP, ED $e$ also makes requests for redundant videos based on pre-fetching decisions $\bm{x}_e\ = [x^k_{e,i}]^{\text{K}_e\times\text{I}}$.
Let $\bm{\lambda}_e\ = [\lambda^k_{e,i}]^{\text{K}_e\times\text{I}}$ denote all video utility values, e.g., the predicted rate to request videos by users, for any ED $e$. The problem of maximizing pre-fetching and caching utility can be formulated by: 
\begin{subequations}
\label{Problem:0}
\begin{align}
    \mathbb{P}_g:\quad &\max_{\bm{x}_e,\forall e} \, \sum_{e\in \mathcal{E}}\sum_{k = 1}^{\text{K}_e}  \sum_{i \in \mathcal{I}} \lambda_{e, i}^k \cdot x^k_{e,i} \\ 
    \mathrm{s.t.}\, 
     & \sum_{i \in \mathcal{I}}  x_{e, i}^k \leq f_{e},   \hspace{55pt}\forall e\in \mathcal{E},\, 1 \leq k \leq \text{K}_e ,\label{problem: P0_con_contentnum}\\
     & x_{e, i}^k \in \{0, 1\},  \hspace{25pt}\forall e\in \mathcal{E},\,\forall i\in \mathcal{I},\,1\leq k\leq \text{K}_e,\\
     & \lambda_{e, i}^k = h_e(i,t^k\ |\ \mathcal{V}_e^k, \bm{\theta}), \hspace{0pt} \forall e\in \mathcal{E},\,\forall i\in \mathcal{I},\,1 \leq k \leq \text{K}_e,
\end{align}
\end{subequations}
where $h_e:\mathcal{I}\times\mathbb{R}^+ \rightarrow \mathbb{R}^+$ can represent any prediction function for utility with model parameters $\bm{\theta}$ and historical viewing records $\mathcal{V}_e^k$ up to time $t^k$ at ED $e$.
Besides, Eq.~\eqref{problem: P0_con_contentnum} restricts the maximum pre-fetching capacity $f_{e}$ of ED $e$.

For traditional video streaming, the CP can collect historical video request records to infer $\bm{\lambda}_e,\, \forall e$, which can be further used to derive optimal solution $\bm{x}_e^*$, for all EDs.  In this process, the CP can exactly infer preferences exposed by EDs. To prevent privacy leakage,  EDs can apply differential privacy (DP) noises to distort pre-fetching decisions $\bm{x}_e$, to hide both users' original video requests and video utility. It is difficult for the CP to infer user privacy from public fetching actions, and hence user privacy is preserved. In the rest of the subsection, we extend $\mathbb{P}_g$ to present the privacy-preserving video pre-fetching problem.

We begin by succinctly introducing DP, avoiding any unnecessary notation.
In problem $\mathbb{P}_g$, the pre-fetching decision variables $\bm{x}$ are primarily determined by the utility $\bm{\lambda}$, which is evaluated by the function $h$ with the parameter $\bm{\theta}$, and the set $\mathcal{V}$ of real request records.
To preserve privacy, DP can be applied to distort the output of utility function $h$ to protect privacy in dataset  $\mathcal{V}$. 

{
\begin{definition}{($\epsilon$-Differential Privacy)} A randomized mechanism $\mathcal{M}$ confirms $\epsilon$-DP, if for any pair of adjacent datasets $\mathcal{V} \simeq\mathcal{V}'$, any tuple of input $(i,t)\in\mathcal{I}\times\mathbb{R}^+$, and any predict function $h$ with its parameters $\bm{\theta}$, it satisfies:
\begin{equation}
    \frac{Pr\{\mathcal{M}(h(i,t\ |\ \mathcal{V}, \bm{\theta})) \in \mathcal{O}\}}{Pr\{\mathcal{M}(h(i,t\ |\ \mathcal{V}', \bm{\theta})) \in \mathcal{O}\}}\leq exp(\epsilon).
\end{equation}
Here, $\epsilon$ is the privacy budget and $\mathcal{O}$ represents the outcome range of mechanism $\mathcal{M}$.
\end{definition}
}

\begin{figure}[t]
\centering
\includegraphics[width=0.95\linewidth]{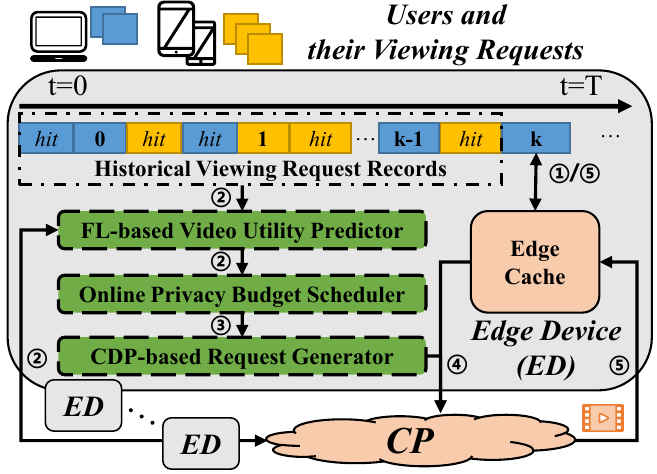}
\caption{The workflow of privacy-preserving video fetching (PPVF) for online video service at EDs.}
\label{fig: steps}
\vspace{-7mm}
\end{figure}

However, in practical online video systems, cardinality $\text{I}$ for $\mathcal{I}$ is a huge number, and users' view preferences can be very skewed, implying that there exists a large number of cold videos with very few user requests~\cite{Zhou2015, Wang2019}. Thereby, directly applying DP noise to distort  utilities for video pre-fetching confronts the following two challenges: 
\begin{enumerate}[label=(\arabic*)]
    \item More privacy budget will be consumed to protect privacy if there are more videos in $\mathcal{I}$. If cardinality $\text{I}$ is a large number, the noises will be excessively large such that the video utility predicted by the function $h$ is valueless.
    \item Considering that the CP can implement collaborative filtering algorithms to infer user privacy, requesting cold videos (with scarce requests) as noises is not effective in preserving privacy since collaborative filtering algorithms can easily remove the noisy influence of cold video requests~\cite{He2017}. 
\end{enumerate}

To tackle these two challenges, we consider adding DP noises to protect pre-fetching privacy by the Exponential Mechanism (EM)~\cite{Zhu2015} with a candidate video set selected in an online manner.
We start with a brief introduction to the EM. The EM is a classical DP mechanism satisfying $\epsilon$-DP, which can be applied to distort the output of utility function $h$, defined as follows.
\begin{definition}{(Exponential Mechanism)}
The exponential mechanism (EM) satisfies $\epsilon$-differential privacy with the following steps: 
(1) specifies a global sensitivity, denoted as $\Delta \lambda$, for a video utility prediction function $h: \mathcal{I}\times\mathbb{R}^+ \rightarrow \mathbb{R}^+$. (2) video $i \in \mathcal{I}$ is selected to request with the probability 
$$P\{i\} \propto \exp\left(\frac{\epsilon \cdot h(i,t\ |\ \mathcal{V}, \bm{\theta})}{2\Delta \lambda}\right).$$
Here, $\epsilon$ represents the privacy budget, $\mathcal{V}$ is the set of request records privately owned by an ED, and $\bm{\theta}$ represents parameters in function $h$.
\end{definition}

Instead of applying the whole video set $\mathcal{I}$, we identify a candidate video set $\mathcal{A}_e^k\subseteq \mathcal{I}$ for ED $e$ to generate redundant requests ED $e$. Here, $k$ is the pre-fetching index for the request missed by the edge cache. 
The whole candidate video set $\mathcal{A}_e = \{\mathcal{A}_e^k\ |\ 1 \leq k \leq \text{K}_e\}$ is obtained by solving the problem \(\mathbb{P}_e\) subjecting to both the privacy budget and pre-fetching capacity constraint. 
In the problem \(\mathbb{P}_e\), the constraint \eqref{problem:Pe_con_epsilon} ensures that the assigned privacy budget cannot exceed the total privacy budget $\xi_{e}$ for each video, and \(f_{e}\) denotes the pre-fetching capability at ED $e$. 
\begin{subequations}
\label{Problem:1}
\begin{align}
    \mathbb{P}_e: \quad& \mathcal{J}_{e}^{*} = \max_{\bm{a}_e} \, \sum_{k=1}^{\text{K}_e}  \sum_{i \in \mathcal{I}} \lambda_{e, i}^k \cdot a^k_{e,i} \\ 
    \mathrm{s.t.}\, 
     & \sum_{k=1}^{\text{K}_e} \epsilon_{e,i} \cdot a_{e,i}^k \leq \xi_{e}, \hspace{87pt} \forall i \in \mathcal{I}, \label{problem:Pe_con_epsilon} \\
     & \sum_{i \in \mathcal{I}}  a_{e, i}^k \leq f_{e},  \hspace{87pt} 1 \leq k \leq \text{K}_e,\label{problem:P_con_contentnum}\\
     & a_{e, i}^k \in \{0, 1\}, \hspace{58pt}\forall i\in \mathcal{I},\,1 \leq k \leq \text{K}_e,\\
     &\lambda_{e, i}^k = h_e(i,t^k\ |\ \mathcal{V}_e^k, \bm{\theta}), \hspace{15pt} \forall i\in \mathcal{I},\,1 \leq k \leq \text{K}_e,\label{EQ:PeLambda}\\
     & \mathcal{A}_e^k =\{i\ |\ a_{e,i}^k=1,\forall i\in\mathcal{I}\},\hspace{30pt}1 \leq k \leq \text{K}_e. \label{EQ:PeA} 
\end{align}
\end{subequations}

Based on problem $\mathbb{P}_e$, we can illustrate the holistic optimization process of PPVF:
\begin{itemize}
\item 
\textbf{\textit{Federated Learning}}: EDs can jointly evaluate video utility, i.e., $\lambda_{e, i}^k = h_e(i,t^k\ |\ \mathcal{V}_e^k, \bm{\theta})$, $\forall\, e,\, i,\, k$, via the federated learning framework. This approach allows EDs to achieve significantly more accurate video utility than those obtained solely from local traces.
\item 
\textbf{\textit{Video Selection and Budget Allocation}}: With evaluated $\bm{\lambda}_{e}^k$, EDs can solve $\mathbb{P}_e$ in an online manner to obtain $\mathcal{A}_e^k$, which is the candidate set of videos to be requested from the remote CP.
\item 
\textbf{\textit{Pre-fetching Request Generation}}: With $\mathcal{A}_e^k$ derived from 
$\mathbb{P}_e$ and privacy budget allocation decisions, EDs can apply the EM to distort their fetching requests with correlated differential privacy (CDP). Then, EDs contact the CP to fetch both viewing videos and redundant videos. 
\end{itemize}

\section{PPVF Framework Design} \label{sec:proposed_alg}

\subsection{PPVF Overview}
To better understand how users, EDs and the CP interact with each other, we present 
the 
workflow of PPVF, as shown in 
Fig.~\ref{fig: steps}. 
Briefly speaking, the life cycle of each request involves five steps. 
\textcircled{1} Upon receiving a video viewing request from 
a 
user, EDs first search for that video in their \textit{edge cache}. 
If the video is cached, EDs 
directly stream it to users with a shorter response latency without any privacy leakage. Otherwise, EDs assemble pre-fetching video requests, along with the viewing video, to fetch redundant videos from the CP.
\textcircled{2} The \textit{utility predictor} evaluates the videos' utility based on the federated learning framework, which ensures that only model parameters \(\bm{\theta}\) are exchanged between EDs and the CP. The process is detailed in Section~\ref{sec: Evaluation Method}. 
\textcircled{3} Subsequently, the \textit{budget scheduler} assesses video utility in conjunction with the privacy budget to curate a video candidate set for the subsequent pre-fetching decision, elaborated in Section~\ref{sec: online algorithm}. 
\textcircled{4} The \textit{request generator} distorts the original utility to generate a pre-fetching decision, which navigates the trade-off between privacy and caching performance by leveraging the EM with the CDP method. 
After combining the pre-fetching decision with the real view video, the final fetching vector is sent to the CP.  
This process will be discussed in Section~\ref{sec: CDP algorithm}. 
\textcircled{5} When the CP returns videos, EDs perform cache replacement based on video utility and forward the viewing video to users.

Note that our main contribution is represented by three modules in green color (long dashed box) in Fig.~\ref{fig: steps}, which will be introduced in the following sections in detail.


\subsection{Threshold-based Online Algorithm}\label{sec: online algorithm}
Directly solving problem $\mathbb{P}_e$ confronts two challenges:
(1) the problem is inherently online due to the dynamic nature of request patterns and video utility, and 
(2) the problem \(\mathbb{P}_e\) can be categorized as an online multiple knapsack problem, which is challenging to solve immediately and irrevocably, even if the utility \(\bm{\lambda}^k_e\) is known.

To solve the challenging online problem $\mathbb{P}_e$, we propose a filtering mechanism~\cite{Li2023,Li2021} that selects a video into $\mathcal{A}_e^k$ if the ratio of this video utility over its excepted privacy budget exceeds a threshold.  
Intuitively speaking, a video of a higher utility will be played by users in the future with a higher probability. Hence, the utility of EDs in caching such videos will be higher.  Meanwhile, considering the limited privacy budget, PPVF only selects videos with the ratio $\lambda_{e,i}^k/\epsilon_{e, i}$ exceeding a threshold.
This threshold is set in accordance with the fraction of the consumed privacy budget.

We can set the threshold for selecting videos as follows. 
Let  $U_e$ and $L_e$ denote the upper and lower bound of the ratio for any video $i$ at ED $e$,  
which means that 
$L_e<\lambda_{e,i}^k/\epsilon_{e, i}<U_e,\ \forall i\in\mathcal{I}, 1 \leq k \leq \text{K}_e$.
With the values of $L_e$ and $U_e$, the threshold function is defined as:
\begin{equation}
\Theta_e(\gamma)=\left\{\begin{array}{cl}
L_e, & 0 \leq \gamma \leq \Gamma_e, \\
\left(\frac{U_e \cdot \exp(1)}{L_e}\right)^{\gamma} \frac{L_e}{\mathrm{e}}, & \Gamma_e<\gamma \leq 1.
\end{array}\right.
\end{equation}
Here, $\gamma\in[0,1]$ denotes the fraction of the privacy budget that has been allocated to a video until the current time, $\Gamma_e= \frac{1}{1+ln(U_e/L_e)}$ is the lowest threshold for assessing the privacy budget proportion $\gamma$.
The intuition of our design is that the selection of a video is more conservative if the consumed privacy budget fraction $\gamma$ of that video is larger. 

The detailed algorithm is presented in Alg.~\ref{alg: online allocation algorithm}.  
Specifically, using the threshold function $\Theta_e(\cdot)$, PPVF randomly selects a video from the set $\mathcal{I}$ and checks whether its $\lambda_{e,i}^k/\epsilon_{e, i}$ is exceeds the threshold $\Theta_e(\gamma_{e,i}^k)$. If so, the video is incorporated into candidate set $\mathcal{A}_e^k$; if not, the video remains unchosen. This stochastic selection process continues until $\mathcal{A}_e^k$ contains $f_e$ videos.
Although Alg.~\ref{alg: online allocation algorithm} is a heuristic-based algorithm, we can theoretically prove that Alg.~\ref{alg: online allocation algorithm} can achieve
an optimal competitive ratio (CR) of $\left(1+\ln(U_e/L_e)\right)$ for any ED $e$.

\begin{theorem} \label{theorem: CR}
Alg.~\ref{alg: online allocation algorithm} has a competitive ratio of $\left(1+\ln(U_e/L_e)\right)$ under rational Assumption~\ref{assumption: tiny cost} for any ED $e$ to allocate the privacy budget.
\end{theorem}
\begin{assumption} \label{assumption: tiny cost}
 Each privacy cost of a pre-fetching video $i$ has a weight much smaller than the total budget of the content, i.e., $\epsilon_{e, i}\ll\xi_{e}$ with respect to any ED $e$.
\end{assumption}
\begin{proof}
We prove Theorem~\ref{theorem: CR} with Assumption~\ref{assumption: tiny cost} in Appendix~\ref{appendix: CR}.
\end{proof}
Considering $U_e=\max_{\forall i,k}\lambda_{e,i}^k/\epsilon_{e,i}$ and $L_e=\min_{\forall i,k}\lambda_{e,i}^k/\epsilon_{e,i}$, we observe that 
CR is solely dependent on $\bm{\lambda}$ and $\bm{\epsilon}$ and is independent of the request quantity (i.e., $K_e$). As the privacy budgets $\bm{\epsilon}$ are specified by EDs for each video, while utility $\bm{\lambda}$ is often generated by upstream utility prediction algorithms, they are both within a specific range. This characteristic ensures a constant-level CR of our algorithm, independent of the total request quantity (i.e., $K_e$), which is a highly appealing property.
Through differentiation of CR with respect to $U_e$ and $L_e$, it becomes evident that a decrease in $U_e$ leads to a reduction in CR, indicating an approach to the optimal offline solution in the worst-case scenario. Similarly, a larger $L_e$ results in a smaller value of CR. Moreover, when $U_e/L_e$ approaches 1, CR approaches 1, indicating that the performance is close to the offline optimal solution.  

\begin{algorithm}[!t]
\caption{Online privacy budget allocation algorithm for ED $e$.}\label{alg: online allocation algorithm}
\KwIn{ The space of all videos $\mathcal{I}$; 
 The total privacy budget for all videos $\xi_{e}$; 
 The pre-fetching capacity $f_e$;
The privacy cost $\bm{\epsilon}_{e}$.
 }
\KwOut{The candidate set $\mathcal{A}_e$ for video pre-fetching.}
 Initialize $k\leftarrow1$, $\bm{\gamma}_{e}^k \leftarrow [0]^{\text{I}}$;\\
\For{$k\leq\text{K}_e$}{
    Initialize $\bm{a}_{e}^k\leftarrow [0]^{\text{I}}$,
    $f^k\leftarrow 0$, $\mathcal{I}^k\leftarrow\mathcal{I}$;\\ 
    Obtain the evaluated video utility $\bm{\lambda}^k_{e}$;\\
    \While{$f^k<f_e$ \textnormal{\textbf{ and }} $\mathcal{I}^k\neq\emptyset$}{
        Select content $i$ randomly from $\mathcal{I}^k$;\\
        $\mathcal{I}^k \leftarrow \mathcal{I}^k- \{i\}$;\\
    \eIf{$\frac{\lambda_{e,i}^k}{ \epsilon_{e,i}}>\Theta(\gamma_{e,i}^k)$ \textnormal{\textbf{ and }}$\epsilon_{e,i}  < (1-\gamma_{e,i}^k)\cdot \xi_{e}$
    }{
 $a_{e,i}^k\leftarrow1$; $\gamma_{e,i}^{k+1} \leftarrow \gamma_{e,i}^{k} + \frac{\epsilon_{e,i}}{\xi_{e}}$; $f^k\leftarrow f^k+1$;\\
        }
        {$a_{e,i}^k\leftarrow 0$; $\gamma_{e,i}^{k+1} \leftarrow \gamma_{e,i}^{k}$; $f^k\leftarrow f^k$;}
    }
    Generate the candidate set $\mathcal{A}_e^k$  with $\bm{a}_{e}^k$ following Eq.~\eqref{EQ:PeA};\\
$k\leftarrow k+1$;\\
}
\end{algorithm}

\subsection{CDP-based Video Pre-fetching}\label{sec: CDP algorithm}

{
Directly requesting videos in  $\mathcal{A}_e^k = \{i\ |\ a_{e,i}^k=1,\forall i\in\mathcal{I}\}$ according to video utility can expose the video utility knowledge to the CP. To preserve privacy, PPVF adopts the EM to randomly select videos based on probability shown in Eq.~\eqref{eq: pre-fetch i probability} and generate the final pre-fetching decision $\bm{x}_e^k$. Specifically, if video \(i\) is selected by the EM, PPVF will set \(x_{e,i}^k = 1\) to pre-fetch that video from the CP. Otherwise, it will be set to \(0\). The probability is given by
\begin{equation}\label{eq: pre-fetch i probability}
P\{\text{video $i$ is chosen from }  \mathcal{\mathcal{A}}_e^k\ |\ \bm{\lambda}_e^k\} \propto  \exp\frac{\epsilon_{e}^k \cdot \lambda_{e,i
}^k}{2\cdot\Delta\lambda_{e,gc}^k}.
\end{equation}
Here, $\epsilon_{e}^k =\frac{1}{f_e}\sum_{i\in\mathcal{A}^k_e} \epsilon_{e,i}$ is the averaged privacy for pre-fetching one redundant video, where $\sum_{i\in\mathcal{A}^k_e} \epsilon_{e,i}$ denotes the total privacy budget assigned by Alg.~\ref{alg: online allocation algorithm}.  
Besides, $\Delta\lambda_{e,gc}^k$ is the global sensitivity at the time of the $k$-th pre-fetching. 
}

In our problem, the calculation of $\Delta\lambda_{e,gc}^k$ is complicated because of the correlation between videos. 
Collaborative filtering algorithms can exploit such correlation information for inferring users' personal interests. To factor in the influence of video correlation, we employ the correlated differential privacy (CDP) for computing sensitivity. 
For ED $e$, we can calculate the correlation between videos $i$ and $j$ for the $k$-th pre-fetching with $\lambda_{e,i(j)}^k$ predicted by utility function $h_e$. 
Suppose that EDs cache three history matrices $\bm{\psi}_e^{k-1} = [\psi_{e,i,j}^{k-1}]^{\text{I}\times\text{I}}$, $\bm{\alpha}_e^{k-1} = [\alpha_{e,i}^{k-1}]^{\text{I}}$, $\bm{\sigma}_e^{k-1} = [\sigma_{e,i}^{k-1}]^{\text{I}}$, where the items can be incrementally updated by 
$\psi_{e,i,j}^{k} = \psi_{e,i,j}^{k-1} + \lambda_{e,i}^{k}\cdot\lambda_{e,j}^{k},\
\alpha_{e,i}^{k} = \alpha_{e,i}^{k-1} + \lambda_{e,i}^{k},\
\sigma_{e,i}^{k} = \sigma_{e,i}^{k-1} + (\lambda_{e,i}^{k})^2,
$ respectively.
Based on Pearson's correlation~\cite{Zhu2015}, the correlation degree between videos $i$ and $j$ can be calculated with these three history matrices as follows:
\begin{equation} \label{eq:correlated degree}
\Psi_{e,i,j}^k=\frac{k\cdot\psi_{e,i,j}^{k}-\alpha_{e,i}^{k} \cdot \alpha_{e,j}^{k}}{\sqrt{k\cdot\sigma_{e,i}^{k} - (\alpha_{e,i}^{k})^2}\cdot\sqrt{k\cdot\sigma_{e,j}^{k} - (\alpha_{e,j}^{k})^2}}.
\end{equation}
Let $\bm{\Psi}_e^k = [\Psi_{e,i,j}^k]^{\text{I}\times\text{I}}$ denote the correlation matrix. 
The sensitivity of our problem can be calculated using the following two definitions.
\begin{definition}{(Correlated Video Sensitivity)} For any given ED $e\in\mathcal{E}$, missing request index $1 \leq k \leq \text{K}_e$ and video $i,j\in\mathcal{A}_e^k$, the correlated video sensitivity $\Delta \lambda_{e,i}^k$ is defined as
\begin{equation}
   \Delta \lambda_{e,i}^k=\sum_{j\in \mathcal{A}^k_e}(\Psi^k_{e,i, j}||h_e(i,t^k|\mathcal{V}^k_e,\bm{\theta})- h_e(i,t^k|\mathcal{V}^{k}_{e,-j},\bm{\theta})||_1)\label{EQ: Delta_i},
\end{equation}
where $\Psi_{e, i, j}^k$ is the correlation degree parameter,
$h_e:\mathcal{I}\times\mathbb{R}^+ \rightarrow \mathbb{R}^+$ is the utility function, $\mathcal{A}^k_e$ is the candidate for video selection, and $\mathcal{V}^k_e$, $\mathcal{V}^{k}_{e,-j}$ are two adjacent datasets different in video $j$.
\end{definition}
Here, the L1 distance measures the effect on utility when deleting records related to video $j$ from $\mathcal{V}_e^k$. 
Parameter $\Psi_{e, i, j}^k$ estimates the correlated degree between videos $i$ and $j$.
Correlated Video Sensitivity combines the effect of correlated records and the correlated degree together.

\begin{definition}{(Correlated Sensitivity~\cite{Zhu2015})} Given all the video sensitivities $\Delta \lambda_{e,i}^k,\,\forall i\in\mathcal{A}_e^k$, the global sensitivity $\Delta \lambda_{e,gc}^k$ for the correlated videos is determined by
\begin{equation}
   \Delta \lambda_{e,gc}^k = \max_{i\in \mathcal{A}_e^k} \Delta \lambda_{e,i}^k.\label{EQ: Delta_gc} 
\end{equation}
\end{definition}
The correlated sensitivity lists all
videos in candidate set $\mathcal{A}_e^k$ responding to utility and selects the maximal video sensitivity as the correlated sensitivity. When videos are independent or weakly correlated, the global sensitivity will 
only slightly increase. 
Particularly, if all videos are independent, the correlated video sensitivity is equal to the global sensitivity, i.e., $\max_{i\in \mathcal{A}^k_e}||h_e(i,t^k|\mathcal{V}^k_e,\bm{\theta})- h_e(i,t^k|\mathcal{V}^{k}_{e,-i},\bm{\theta})||_1$.
{
Additionally, given the historical matrices $ \bm{\psi}_e^k$, $ \bm{\alpha}_e^k$, and $ \bm{\sigma}_e^k$, which can be incrementally updated prior to the $k$-th pre-fetching, the correlated degree $ \Psi_{e,i,j}^k$ can be efficiently determined in $ \text{O}(1)$, as per Eq.~\eqref{eq:correlated degree}. Consequently, the computational complexity to obtain sensitivity $ \Delta\lambda^k_{e,i}$ of video $i$ is $ \text{O}(f_e)$ with the evaluated utility vector $ \bm{\lambda}^k_e$ by function $h_e$ and the correlation matrix $\bm{\Psi}_{e}^k$.
Finally, the total computational complexity for deriving the correlated sensitivity $\Delta \lambda^k_{e,gc}$ is $ \text{O}(f_e^2)$ with the utility vector $\bm{\lambda}^k_e$ and the correlation matrix $ \bm{\Psi}_{e}^k$.
}

The detailed algorithm to generate redundant pre-fetching video requests is presented in Alg.~\ref{alg: online pre-fetching algorithm}. 
It can be proved that Alg.~\ref{alg: online pre-fetching algorithm} guarantees a $\sum_{i\in\mathcal{A}_e^k}\epsilon_{e,i}$-DP for the $k$-th pre-fetching decision at ED $e$, where $\sum_{i\in\mathcal{A}^k_e} \epsilon_{e,i}$ denotes the total privacy budget assigned by Alg.~\ref{alg: online allocation algorithm}.
{
The proof can be directly deduced by considering
the properties of the EM and the Composition Theorem~\cite{McSherry2009}.
The EM facilitates the selection of one video from the candidate set for pre-fetching while ensuring $\epsilon_e^k$-DP compliance. Based on the Composition Theorem~\cite{McSherry2009}, Alg.~\ref{alg: online allocation algorithm} makes a maximum $f_e$ times of selections, adhering to $f_e \cdot \epsilon_e^k$-DP, which is equivalent to $\sum_{i\in\mathcal{A}^k_e} \epsilon_{e,i}$-DP.
}
{
\begin{remark}
    In a nutshell, the superiority of PPVF for optimally balancing privacy preservation and efficiency is attributed to the following two advantages. Firstly, rather than blindly distorting requests for all videos, PPVF can reduce the consumption of the privacy budget by only distorting requests for a subset of selected candidate videos.  
    Secondly, by considering correlation in video requests, PPVF can more accurately calibrate the noise scale by using  CDP, which can avoid setting over-large noise scales for videos.   
\end{remark}
}

\begin{algorithm}[t]
\caption{Online privacy-preserving videos pre-fetching algorithm for ED $e$.}\label{alg: online pre-fetching algorithm}
\KwIn{ 
Pre-fetching index $k$; 
Video utility $\bm{\lambda}^k_{e}$;
Allocated privacy budget $\bm{\epsilon}_{e}$; 
Pre-fetching capacity $f_e$. 
 }
\KwOut{The pre-fetching decision $\bm{x}_e$.}
Obtain candidate set $\mathcal{A}_e^k$ for the $k$-th pre-fetching by Alg.~\ref{alg: online allocation algorithm};\\
Incrementally update \( \bm{\psi}_e^k \), \( \bm{\alpha}_e^k \), and \( \bm{\sigma}_e^k \) with video utility $\bm{\lambda}^k_{e}$;\\
Calculate $\bm{\Psi}_{e}^k$ with matrices \( \bm{\psi}_e^k \), \( \bm{\alpha}_e^k \), and \( \bm{\sigma}_e^k \) by Eq.~\eqref{eq:correlated degree};\\
Obtain $\Delta \lambda_{e,gc}^k$ with $\bm{\lambda}^k_{e}$, $\bm{\Psi}_{e}^k$ and $\mathcal{A}_e^k$ following Eqs.~\eqref{EQ: Delta_i}-\eqref{EQ: Delta_gc};\\
$\bm{x}_{e}^k\leftarrow [0]^{\text{I}}$, $f^k\leftarrow 0$;\\
\While{$f^k<f_e$}{
  Select pre-fetching video $i$ from $\mathcal{A}_e^k$ based on the probability shown in Eq.~\eqref{eq: pre-fetch i probability};\\
  $x_{e,i}^k = 1$, $f^k\leftarrow f^k + 1$;\\
}
\Return Pre-fetching decision $\bm{x}_{e}^k$.
\end{algorithm}

\section{Online Video Utility Prediction} \label{sec: Evaluation Method}
In this section, we shift our focus to the discussion on evaluating video utility, i.e., $\lambda_{e, i}^k = h_e(i,t^k\ |\ \mathcal{V}_e^k, \bm{\theta}),\forall e,\,i,\,k$, by federated learning. Note that FL is a privacy-preserving framework for training machine learning models. We resort to point process-based models, i.e., Mutual-Exciting Process (MEP)~\cite{Daley2008, Hawkes1971}, to illustrate how PPVF works. 
Note that MEP is employed here because it has been widely used in~\cite{Shi2021, Zhang2022, Rizoiu2017} for predicting video utility in traditional online video caching problems. 
It is not difficult to replace MEP with a new model for video utility prediction in PPVF. 
In this section, we focus on how to modify it to fit in PPVF.  
 
\subsection{Intensity and Likelihood Function}

The core of a point process lies in its intensity function, denoting the occurrence probability of an event within a tiny time window $[t, t+\rm{d}t)$~\cite{Rizoiu2017}. By abusing notations a little bit, an intensity function can be defined by $h(\iota,t)\text{d}t = P\{ \Omega \, |\, \mathcal{V}(t)\} = \text{E}(\text{d} N(\iota,t)|\mathcal{V})$, where $N(\iota,t)$ is the count function and $\text{E}(\text{d}N(\iota,t)\, |\, \mathcal{V})$ represents the expected count of occurrences of event $\Omega$ with type $\iota$ in the time window $[t, t+\text{d}t)$ based on the historical event set $\mathcal{V}$~\cite{Rizoiu2017}.

In PPVF, the historical event set $\mathcal{V}$ corresponds to the historical record set of viewing requests. We can create an intensity function for a particular video $i$ (i.e., event type $\iota$) indicating the expected request rate from users for that video, which is regarded as the utility of video $i$ for caching. 
Recall that the set $\mathcal{T}_{e,i}^{0,t^k}$ represents the timestamps corresponding to requests of video $i$ in local viewing records $\mathcal{V}_e^k$ at ED $e$ within the time interval $[0,t^k)$.
We can create the intensity function for video $i$ and time $t^k$ at ED $e$ as: 
\begin{equation}
\begin{aligned}
h_e(i, t^k \,|\, \mathcal{V}_e^k,\bm{\beta},\bm{\omega})&= \beta_{i}+ \sum_{ j\in\mathcal{I}}\omega_{ij} \sum_{\tau \in \mathcal{T}_{e, j}^{0,t^k}} \phi(t^k-\tau),
\label{EQ: lambda_{e,i}^k}
\end{aligned}
\end{equation}
where $\bm{\omega} = [\omega_{i,j}]^{\text{I}\times\text{I}},\omega_{i,j}\in\mathbb{R}^+$ denotes the influencing parameter matrix among all videos, while $\bm{\beta} = [\beta_i]^{\text{I}},\beta_{i}\in\mathbb{R}^+$ is the bias parameter vector of the intensity functions. 
Specifically, $\phi(\cdot)$ is defined as $\phi(t) = \exp(-\delta \cdot t)$, where exponential decreasing kernel functions are adopted to gauge the influence of historical events for point process models~\cite{Rizoiu2017, Hawkes1971, Daley2008}.  Here, $\delta> 0$ serves as a hyper-parameter of the influence attenuation coefficient.{
\begin{remark}
 The intuition of Eq.~\eqref{EQ: lambda_{e,i}^k} is that users' video requests at different time points are correlated, where a more recent historical video request would contribute a higher request rate to its relevant video. 
 The extent can be captured by influencing parameters and kernel functions in the point process model to predict future request rates.     
\end{remark}}
To make our presentation concise, $h_e(i,t^k)$ is used to represent $ h_e(i, t^k \,|\, \mathcal{V}_e^k,\bm{\beta},\bm{\omega})$ hereafter if the meaning is clear. 
The parameter space of $\omega_{i,j}$ in Eq.~\eqref{EQ: lambda_{e,i}^k} is $\text{O}(\text{I}^2)$ which is prohibitive for solving directly. The parameter space can be reduced by Singular Value Decomposition (SVD)~\cite{Shi2021}. Given that $\omega_{i,j}$ represents how much video $j$ influences video $i$, it can be decomposed as the product of $\omega_{i,j} = \bm{p}_i\cdot \bm{q}_j^\intercal$. 
Here, $\bm{p}_i$ and $\bm{q}_j$ are latent vectors with dimension $\text{D}\ll\text{I}$.
Hence, we can significantly shrink the dimension space of  $\bm{\omega}$ to avoid overfitting.
Specifically, the parameter dimensions can be condensed from \( \text{I} \times \text{I} \) to \( 2 \times \text{I} \times \text{D} \), where $\text{D}\ll\text{I}$.
Consequently, the utility $\lambda_{e, i}^k,\forall e,i,k$ can be obtained by the revised form in Eq.~\eqref{EQ: lambda_{e,i}^k com}.
\begin{equation}
\begin{aligned}
\lambda_{e,i}^k=\hat{h}_e(i, t^k) &= \beta_{i}+ \sum_{ j\in\mathcal{I}}\bm{p}_i\cdot \bm{q}_j^\intercal \sum_{\tau \in \mathcal{T}_{e, j}^{0,t^k}} \phi(t^k-\tau).
\label{EQ: lambda_{e,i}^k com}
\end{aligned}
\end{equation}

Next, we can use the maximum likelihood estimation (MLE)~\cite{Zhang2022} to optimize all parameters denoted by $\bm{\theta} \defeq \{\bm{\beta},\bm{p},\bm{q}\}$ in Eq.~\eqref{EQ: lambda_{e,i}^k com}.
With local timestamp set $\mathcal{T}_{e,i}$ for video $i$ in time $[0,T)$ at ED $e$, the local log-likelihood function is derived as:
\begin{equation}
\begin{split}
    ll_e \left( \, \bm{\theta}\, | \,\mathcal{V}_e\,\right) =   \sum_{i\in \mathcal{I}} \sum_{ \tau \in \mathcal{T}_{e,i}} \log \hat{h}_e(i,\tau) - \int_{0}^{T} \hat{h}_e(i,t) \mathrm{d} t. 
\label{EQ:logL}
\end{split}
\end{equation}
Here, $\mathcal{V}_e$ represents the whole private dataset at ED $e$ to generate the time timestamp $\mathcal{T}_{e,i}$ for any $i\in\mathcal{I}$.
The detailed derivation can be found in Appendix~\ref{appendix:log-likelihood}.
To preserve privacy, each ED $e$ should locally maximize Eq.~\eqref{EQ:logL}. 
However, the estimation accuracy will be inferior because request records owned by each ED can be very scarce. 
Moreover, given the dynamic nature of video popularity, parameter estimation cannot be solved by one-time training. Continuous online learning is necessary to closely track the changes in video request patterns.  
To address these challenges, we propose an FL-based online parameter estimation algorithm, and the CP can coordinate the training process by collecting, aggregating, and distributing model parameters. 




\subsection{FL-based Online Parameter Estimation}

\subsubsection{\textbf{Local Online Log-Likelihood Function for EDs}}
In practical video systems, user requests are generated 
online, 
which can make the computation complexity of $\lambda_{e,i}^k$ very heavy. To alleviate computation overhead, we simplify Eq.~\eqref{EQ: lambda_{e,i}^k com}  by removing distant historical events without compromising the accuracy of utility prediction. 
In Eq.~\eqref{EQ: lambda_{e,i}^k com},
the kernel $\phi(t-\tau)$ represents the influence of the request at time $\tau$ on video utility, where $t$ is the current time. 
If $t-\tau\gg 1$, it implies that  $\phi(t-\tau)\approx 0$ and the influence of the request at the past time $\tau$ on video utility prediction is negligible. 
Thus, we set a threshold $\phi_{th}$ to eliminate these distant records from computation. 
At time $t$, all records before time $t+\frac{\log\phi_{th}}{\delta}$ will be ignored. In this way, we can significantly reduce computation overhead. 


Let $\Delta t= -\frac{\ln{\phi_{th}}}{\delta}$. At time $t_\theta$, where $t_{\theta}\in\mathcal{Q}$ is the timestamp to update model parameters, we only consider records in the period $[t_{\theta}-\Delta t, t_{\theta})$ in an online manner.
\begin{equation}
    \begin{split}
       \Hat{ll}_e \left( \, \bm{\theta} \,|\,\mathcal{V}_e \right) &=    \sum_{i\in \mathcal{I}} \left(\sum_{ \tau \in \mathcal{T}_{e,i}^{t_{\theta}-\Delta t,t_{\theta}}} \log \hat{h}_e(i,\tau)\right. \\
        &\hspace{80pt} - \left.\int_{t_{\theta}-\Delta t}^{t_{\theta}} \hat{h}_e(i,t) \mathrm{d} t \right).
         \label{EQ:logL online}
    \end{split}
\end{equation}
Notably, unlike the approach in~\cite{Zhang2022}, we only truncate the online update interval for the log-likelihood calculation while preserving the influence of all historical events in the intensity function. This choice aligns with our ability to incrementally compute the impact of historical records in Eq.~\eqref{EQ: lambda_{e,i}^k com}, Eq.~\eqref{EQ:logL online}, avoiding excessive computational complexity.
For a detailed discussion on the incremental computation of the point process, please consult~\cite{Shi2021}.

With a direct mathematical derivation, the partial derivatives of each parameter respected to Eq.~\eqref{EQ:logL online} can be derived as follows:

\begin{align}
\begin{split}
\frac{\partial \Hat{ll}_e(\bm{\theta}\,|\,\mathcal{V}_e)}{\partial \beta_{i}}&= \sum_{\tau\in\mathcal{T}_{e,i}^{t_{\theta}-\Delta t,t_{\theta}}}\frac{1}{\hat{h}_{e}(i,\tau)}-\Delta t, \label{eq: ll dev betai}
\end{split} 
\\ 
\begin{split}
\frac{\partial \Hat{ll}_e(\bm{\theta}\,|\,\mathcal{V}_e)}{\partial \bm{p}_{i}} &= \sum_{\tau\in\mathcal{T}_{e,i}^{t_{\theta}-\Delta t,t_{\theta}}}\frac{\sum_{j\in \mathcal{I} } \bm{q}_j^\intercal\sum_{\tau'\in\mathcal{T}_{e,j}^{0,\tau}} \phi(\tau-\tau')}{\hat{h}_{e}(i,\tau)}\\
&-\sum_{j'\in \mathcal{I} }\bm{q}_{j'}^\intercal\left(\sum_{\tau''\in\mathcal{T}_{e,j'}^{0,t_{\theta}-\Delta t}}\int_{t_{\theta}-\Delta t-\tau'' }^{t_{\theta}-\tau''} \phi(t) \mathrm{d} t \right. \\
&-\left.\sum_{\tau'''\in\mathcal{T}_{e,j'}^{t_{\theta}-\Delta t, t_{\theta}}}\int_{0}^{t_{\theta}-\tau'''} \phi(t) \mathrm{d} t\right),\label{eq: ll dev p i}
\end{split}  
\\
\begin{split}
\frac{\partial \Hat{ll}_e(\bm{\theta}\,|\,\mathcal{V}_e)}{\partial \bm{q}_{j}} &= \sum_{i\in\mathcal{I}}\bm{p}_i\left(\sum_{\tau\in\mathcal{T}_{e,i}^{t_{\theta}-\Delta t,t_{\theta}}}\frac{ \sum_{\tau'\in\mathcal{T}_{e,j}^{0,\tau}} \phi(\tau-\tau')}{\hat{h}_{e}(i,\tau)}\right.\\
&-\sum_{\tau''\in\mathcal{T}_{e,j}^{0,t_{\theta}-\Delta t}}\int_{t_{\theta}-\Delta t-\tau'' }^{t_{\theta}-\tau''} \phi(t) \mathrm{d} t \\
&-\left.\sum_{\tau'''\in\mathcal{T}_{e,j}^{t_{\theta}-\Delta t, t_{\theta}}}\int_{0}^{t_{\theta}-\tau'''} \phi(t) \mathrm{d} t\right).\label{eq: ll dev q i}
\end{split}
\end{align}

Upon computing the local likelihood value \( \Hat{ll}_e(\bm{\theta}\,|\,\mathcal{V}_e) \) and gradient values \( \frac{\partial \Hat{ll}_e(\bm{\theta}\,|\,\mathcal{V}_e)}{\partial \beta{i}} \), \( \frac{\partial \Hat{ll}_e(\bm{\theta}\,|\,\mathcal{V}_e)}{\partial \bm{p}_{i}} \), and \( \frac{\partial \Hat{ll}_e(\bm{\theta}\,|\,\mathcal{V}_e)}{\partial \bm{q}_{j}} \) using local historical viewing records from time \( [t_{\theta}-\Delta t,t_{\theta}) \), the EDs transmit only these non-sensitive gradients and likelihood values to the CP for global estimation. 

\subsubsection{\textbf{Global Log-Likelihood Function for CP}}
For a more precise understanding of the global estimation, we present the global likelihood function as follows:
\begin{equation}
\begin{split}
\min_{\bm{\theta}}\ L = - \sum_{e\in\mathcal{E}}\Hat{ll}_e &\left( \, \bm{\theta} \,|\,\mathcal{V}_e \right)+\frac{\rho_\beta}{2}\lVert \bm{\beta}\rVert_2^2+ \frac{\rho_p}{2}\lVert \bm{p}\rVert_2^2 + \frac{\rho_q}{2}\lVert \bm{q}\rVert_2^2 ,
\\
&\text { s.t. } \quad \bm{\beta},  \bm{q}, \bm{p} \in \mathbb{R}^+,
\label{eq: global loss}
\end{split} 
\end{equation}
where \( \rho_{\beta},\rho_{q},\rho_{p} > 0 \) denote regularization parameters. 
Here, instead of direct computation by the server using private records, all EDs upload the local likelihood function value \( \Hat{ll}_e\left( \bm{\theta} \,|\,\mathcal{V}_e\right) \). Furthermore, for any \( \theta \in \bm{\theta} \), the CP can aggregate the gradient of \( \theta \) separately, drawing upon the gradient \( \frac{\partial \Hat{ll}_e(\bm{\theta}\,|\,\mathcal{V}_e)}{\partial \theta} \) provided by the EDs. Subsequently, the partial derivative from Eq.~\eqref{eq: global loss}, when aggregated in CP, aligns with
\begin{equation}
    \frac{\partial L}{\partial \theta}=\rho_\theta\theta + \sum_{e\in\mathcal{E}} \frac{\partial \Hat{ll}_e(\bm{\theta}\,|\,\mathcal{V}_e)}{\partial \theta}\label{Eq: global partial}.
\end{equation}
Let \( \theta^{(n)} \) denote the parameters trained for \( n \) iterations.
Subsequently, the update rule for \( \theta \) is:
\begin{equation}
\theta^{(n+1)} \leftarrow \theta^{(n)}+\eta \left(- \frac{\partial \, L}{\partial \, \theta^{(n)}} + \rho_{\theta} \theta^{(n)}\right).
\end{equation}
Here, \( \rho_{\theta} \) represents the regularization parameter specific to \( \theta \), while \( \eta \) signifies the learning rate determined by the selected gradient descent algorithm. Upon completing a round of parameter updates, the updated parameters $\bm{\theta}^{(n)}$ are disseminated to each ED for the subsequent iteration.

\subsubsection{\textbf{FL-based Execution}}

In our FL framework, EDs
compute local likelihood plus gradient values, and subsequently expose these parameters (i.e., $\bm{\theta}$) to the parameter server (perhaps maintained by the CP) for parameter aggregation. By interacting with EDs, the CP is responsible for collecting, aggregating, updating, and then disseminating model parameters to EDs. This approach is crafted to optimize the model without 
sharing 
raw historical viewing records from EDs, and thus preserves privacy. 

It can be completed by iteratively conducting the following two operations on EDs and the CP: 
\begin{itemize}
    \item \textit{EDs' role in FL}:  EDs calculate the local likelihood function and gradients using recent timestamp sets \( \cup_{i\in\mathcal{I}}\mathcal{T}^{t_{\theta}-\Delta t,t_{\theta}}_{e,i} \) related to historical record set, in conjunction with model parameters \( \bm{\theta}^{(n)} \) from the \( n \)-th iterations. Then, EDs transmit these gradients and likelihood values to the parameter server. Therefore, user privacy is preserved since the original data will not be exposed.
    
    \item \textit{CP's role in FL}: Once the CP receives computations from EDs, it aggregates likelihood functions and gradient values for all EDs. This consolidated result underpins the global gradient update. Following this update, the CP distributes the revised model parameters \( \bm{\theta}^{(n+1)} \) to EDs. 
\end{itemize}
The details can be found in Alg.~\ref{alg:Online train} in Appendix~\ref{appendix: partial derivatives}.

\section{Performance Evaluation}\label{sec: Experiment}

In this section, we conduct trace-driven experiments to evaluate PPVF using a real viewing dataset in Tencent Video. 
We seek to answer the following three questions:
(1) How well do PPVF's components of \textit{budget scheduler} and \textit{request generator} perform in distorting the private information in user profiles (Section~\ref{SEC: Privacy Protection RESULTS JS})?
(2) How well does PPVF work to protect users' interests against the powerful recommendation system (Section~\ref{SEC: Privacy Protection RESULTS RHR})? 
(3) How well can the PPVF framework adapt to traffic changes and improve edge caching performance compared to fixed experts and SOTA learning-based approaches (Section~\ref{sec: caching results})? 

To facilitate the peer review, we also 
release 
the source code of our system 
PPVF\footnote{https://github.com/zhangxzh9/PPVF-MAINCODE} and the dataset\footnote{https://github.com/zhangxzh9/PPVF-DATASET}. We now discuss the methodology and setup of our evaluations.

\subsection{Dataset and Settings}\label{SEC:Dataset and Settings}
Given the large scale of Tencent Video Datasets, we randomly sample a small subset of $10,000$ users drawn from the upper echelon of active users (i.e., the set of $20\%$ users with the largest interactive viewing records) in the origin public dataset~\cite{Zhang2022}. The new dataset consists of $933,541$ video viewing requests for $10,373$ unique videos, all within a specific city over 30 days.
Following~\cite{Shi2021}, we evenly group these users into $25$ fixed groups to replicate a real online video system aided by $25$ EDs during the whole 30 days. It is important to note that this number set is simulation-based, and this setting can be tuned by the customized strategy 
that is 
adaptable to various scenarios 
and meets 
different user privacy requirements. The experimental results also demonstrate the robustness of our framework at different levels of EDs. 
As such, each request record in the dataset is collated by the metadata $(e, u, i, \tau)$.

Similar to~\cite{Zhang2022},  the time interval in our experiments is quantized at $1$ hour for all requests, i.e., the requests arrived at the same hour have the same timestamp $\tau$.
It is 
important to emphasize that
the pre-fetching video's timing is not tied to a specific time slot. 
If a request is not met at the edge, EDs must promptly retrieve the video from 
CP using the pre-fetching algorithm. 
Additionally, we split the dataset into two date-based subsets. The initial subset, containing requests with time $0\leq \tau < 240$ hours, is used to initialize the system. The subsequent subset with requests over the next $20$ days (i.e., $240 \leq \tau < 720$ hours) functions as the test period. 

Other experimental setups are described according to different tasks:
(1) 
\emph{\textbf{Point process and FL}}: 
Following~\cite{Shi2021}, the decay parameter and the dimension of the latent vector are designated as $\delta = 0.01$ and $D=10$, respectively. In alignment with~\cite{Zhang2022}, all model parameters (i.e., $\bm{\beta}$, $\bm{p}$, and $\bm{q}$) are initialized at $1.0$. 
For online FL-based parameter estimation, the maximum iteration count is $20$ and the truncated threshold is set to $\phi_{th} = e^{-0.48}$. Therefore, the interval $t_{\theta}$ to update $\bm{\beta}$, $\bm{p}$, and $\bm{q}$ is $2$ days ($48$ hours), which is the same as that setting in~\cite{Zhang2022}. Some detailed experiments are conducted to study the influence of the online updated interval of the FL framework.
(2) 
\emph{\textbf{Edge pre-fetching and caching}}: 
For experimental consistency, the privacy cost ($\epsilon_{e,i},\forall e,\, i$) is uniformly set to $1$ for each video's pre-fetching requests~\cite{Pan2023}. Additionally, we standardize the allocation of the privacy budget, pre-fetching capacity, and caching capacity across all EDs with values set at $\xi_e = 15$, $f_e = 4$, and $c_e = 1\%$, respectively. In specific experiments, one of these 
parameters might be varied to assess its impact, with the other two 
being constant.

\subsection{Baselines} 
We compare PPVF with three types of baselines. The first type includes privacy-preserving video fetching algorithms, 
while 
the second and third types are video caching algorithms 
that do not consider 
privacy leakage. 
Privacy-preserving caching algorithms include:
(1) 
\emph{\textbf{SAGE}} \cite{Lecuyer2019}, which pre-fetches videos with randomly assigned privacy budgets until the privacy budgets reach the maximum constraints;
(2) 
\emph{\textbf{BESTFIT}}, which allocates the privacy budget to pre-fetch videos with the highest utility until the privacy budgets reach the maximum constraints.
Note that these two baselines are only designed for privacy allocation without designing a video utility predictor. For a fair comparison, we implement our video utility predictor in SAGE and BESTFIT for caching.

To further demonstrate the superiority of our utility predictor, we 
replicate 
two advanced caching utility prediction methods at the edge for comparison. These algorithms are introduced as follows:
(3) 
\emph{\textbf{MAV}~\cite{zhang2023crvr}}, which caches the videos at the edge nodes considering the strength of user requests in the future round within the dynamic Stackberg game. The caching utility is calculated by the moving average value (MAV) method, and the weight of MAV is set as $0.9$ based on \cite{zhang2023crvr};
(4) 
\emph{\textbf{HRS}~\cite{Zhang2022}}, which serves as a video popularity prediction model designed for the edge server, employing a fusion of three distinct point process models. All parameter configurations within this baseline align with the defaults specified in~\cite{Zhang2022}. 
It is worth mentioning that these two baselines are mainly designed to improve edge caching efficiency with utility (e.g., popularity) prediction. Both of them 
overlook the privacy of users 
exposed by pre-fetching requests. Therefore, we only replace our utility predictor module with MAV and HRS and keep the other system components unchanged. 

We also compare PPVF with the following
two eviction caching algorithms, in which EDs only fetch videos watched by users when the cache is missed. These algorithms include:
(5) \emph{\textbf{LRU} (Least Recently Used)}, which replaces the video that has not received any request for the longest time with a newly requested video;
(6) \emph{\textbf{LFU} (Least Frequently Used)}, which replaces the video that has been requested in the least number of times with a newly requested video. 
These two conventional caching algorithms are extensively employed both in industry and academia, making them suitable benchmarks for comparing caching performance.

\subsection{Metrics} 
To evaluate PPVF, 
we employ three metrics to evaluate both privacy protection and system efficiency. More specifically, we adopt the following metrics in our experiments:
\begin{enumerate}
    \item 
    \emph{\textbf{JS} (Jaccard Similarity)} measures the averaged similarity between users' real profiles and profiles exposed by their ED for video fetching. Each profile is represented by a vector of dimension $\text{I}$, where each element indicates whether a video has been requested by a user or ED during the entire testing period. A lower similarity is more desirable, implying stronger privacy protection. 
\item 
\emph{\textbf{RHR} (Recommendation Hit Rate) \textbf{Degradation}}, which calculates the averaged degradation of RHR among all users when using a recommendation algorithm to recommend videos for users based on their original profiles and noisy profiles exposed by EDs. A larger degradation of RHR implies stronger privacy protection.
A popular collaborative filtering recommendation algorithm~\cite{He2017}, NCF, is implemented with the same settings in \cite{He2017} as the adversary in our experiments.  
\item 
\emph{\textbf{CHR} (Cache Hit Ratio)}, which is defined as the number of video hits at all EDs divided by the total number of original video requests from users over the entire test period. CHR is employed to evaluate the caching system efficiency. 
\end{enumerate}

\begin{figure*}[!t]
  \centering
  \begin{minipage}[t]{0.24\textwidth}
    \centering
\begin{subfigure}{\linewidth}
    \includegraphics[width=\linewidth]{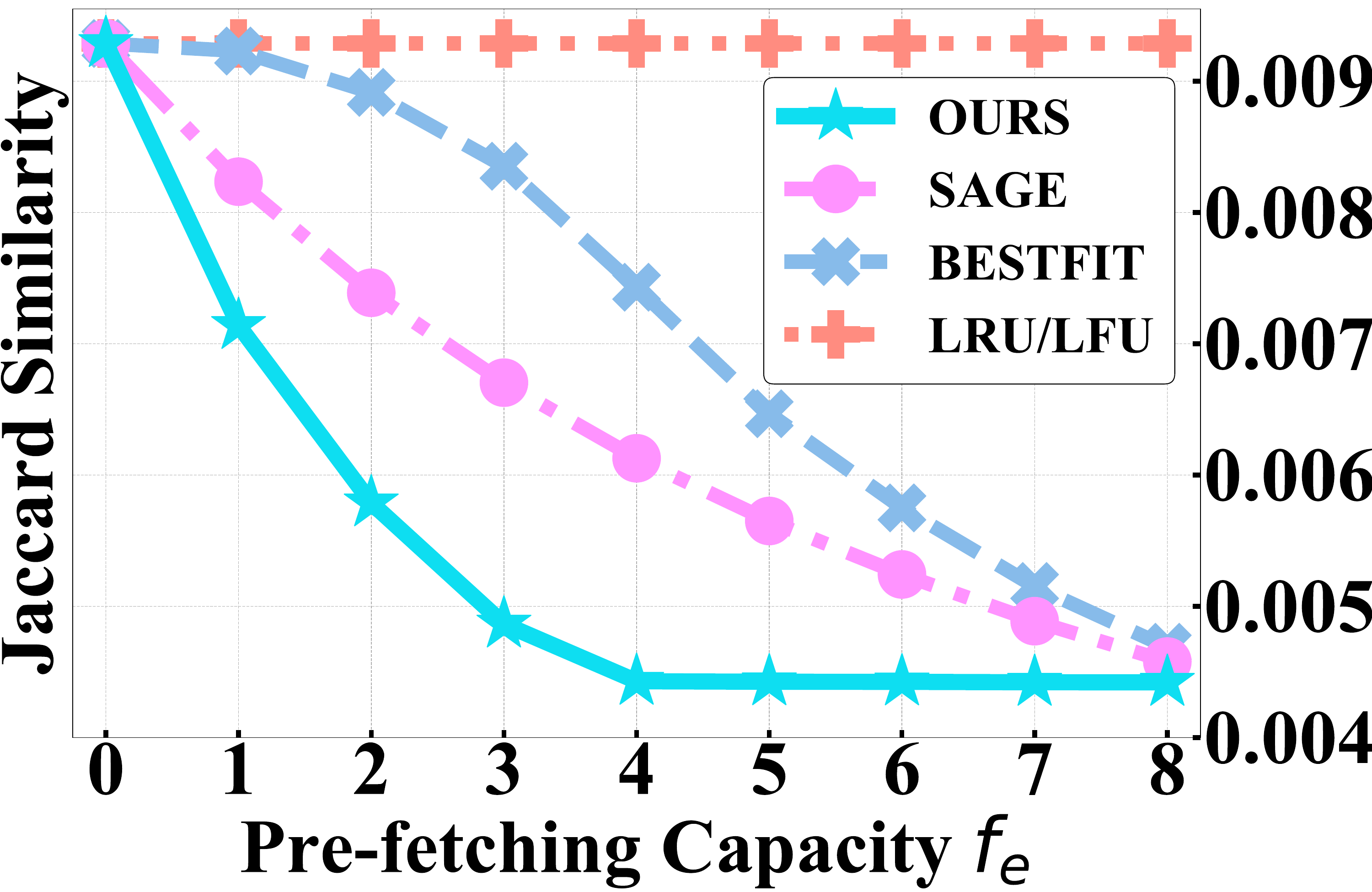}
    \caption{The average JS between users' origin and distorted profile when varying pre-fetching capacity $f_e$ at EDs.}
    \label{fig:fe_change_jaccard}
\end{subfigure}
  \end{minipage}
  \hfill
  \begin{minipage}[t]{0.24\textwidth}
\begin{subfigure}{\linewidth}
    \centering
    \includegraphics[width=\linewidth]{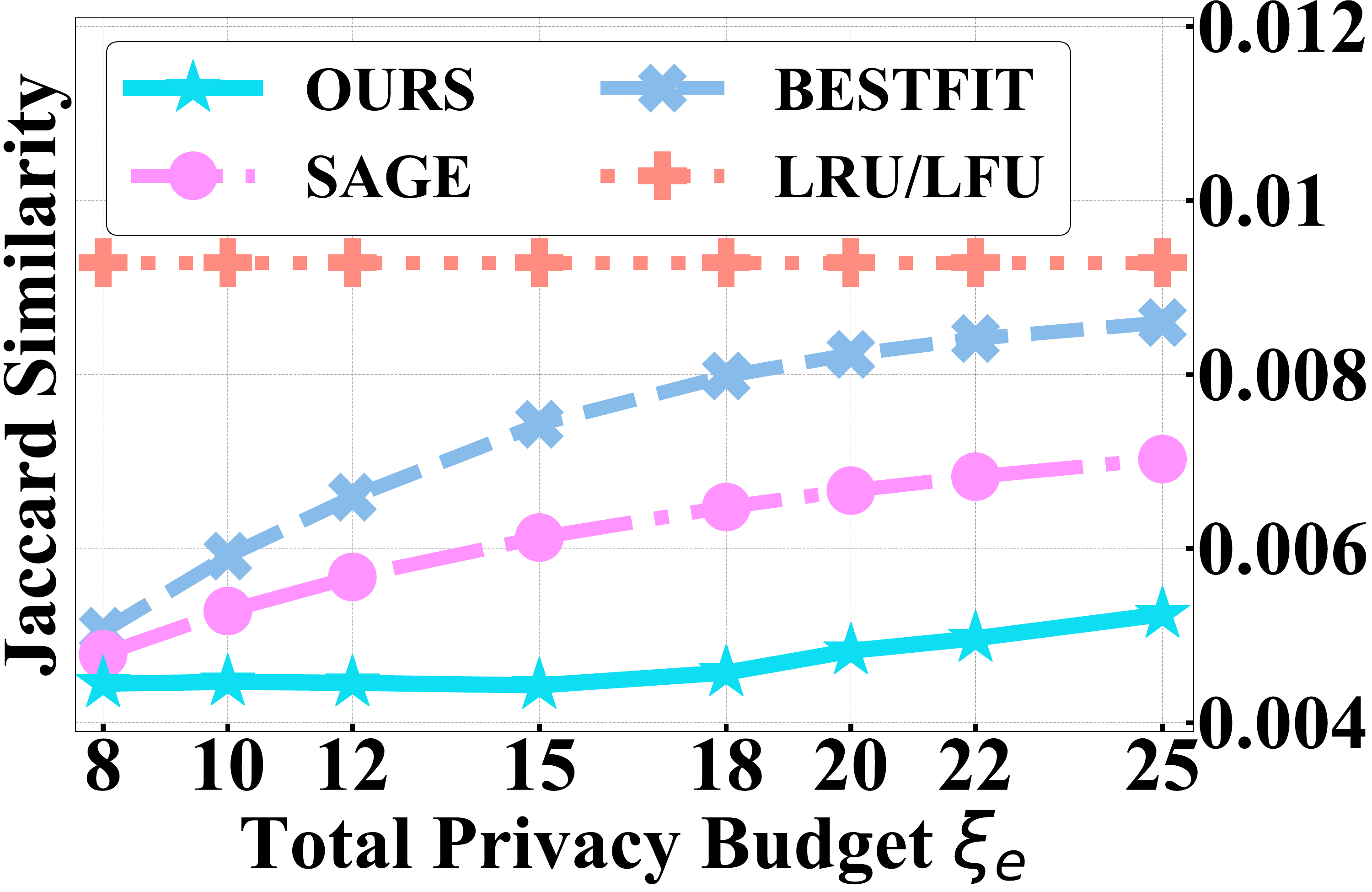}
    \caption{The average JS between users' origin and distorted profile when varying total privacy budget $\xi_e$ at EDs.}
    \label{fig:xi_change_jaccard}
\end{subfigure}
  \end{minipage}
  \hfill
  \begin{minipage}[t]{0.24\textwidth}
    \centering
\begin{subfigure}{\linewidth}
    \includegraphics[width=\linewidth]{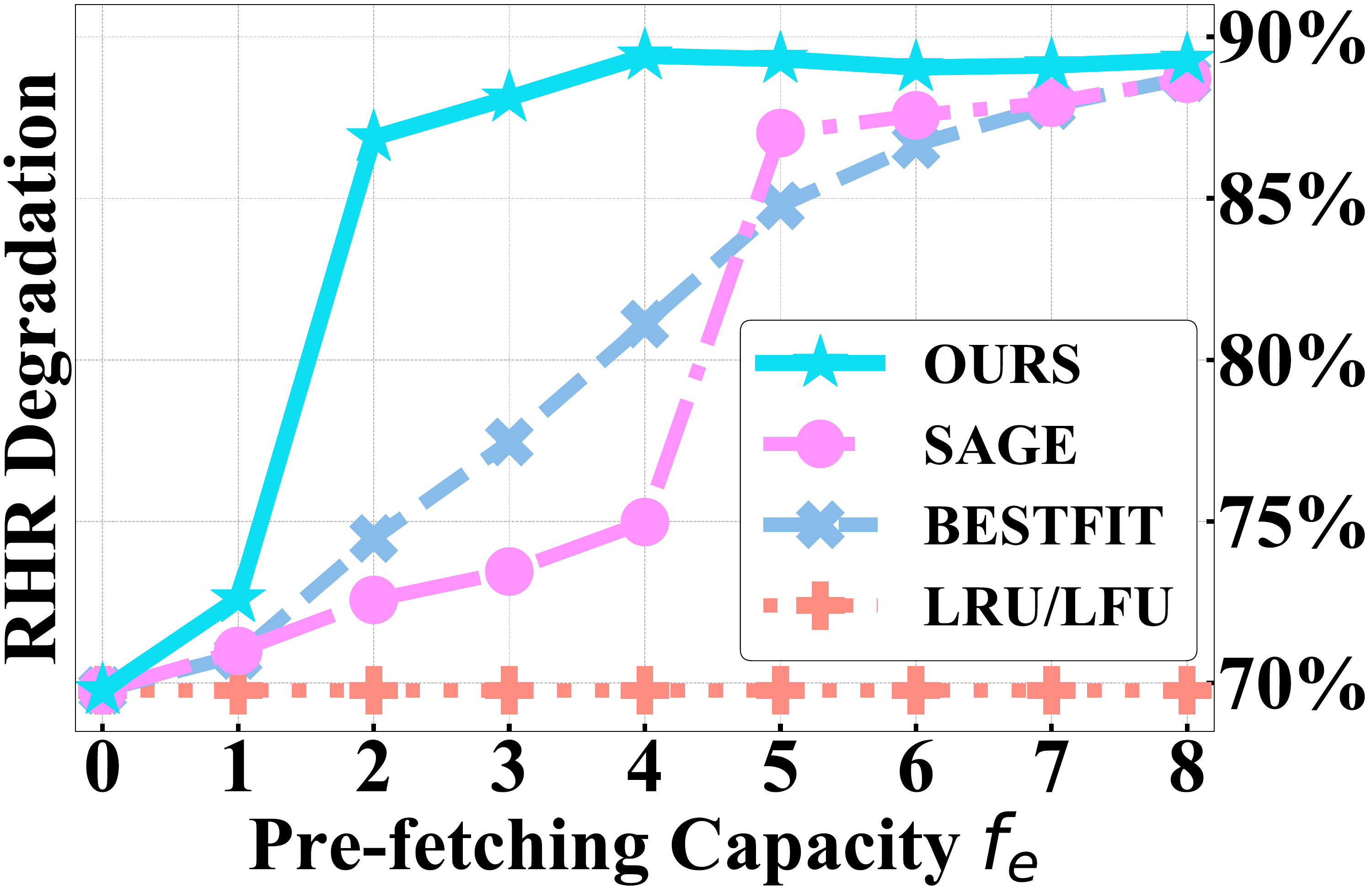}
    \caption{The average decline in RHR between users' origin and distorted profile when varying pre-fetching capacity $f_e$ at EDs.}
    \label{fig:fe_change_DRHR}
\end{subfigure}
  \end{minipage}
    \hfill
  \begin{minipage}[t]{0.24\textwidth}
    \centering
\begin{subfigure}{\linewidth}
\includegraphics[width=\linewidth]{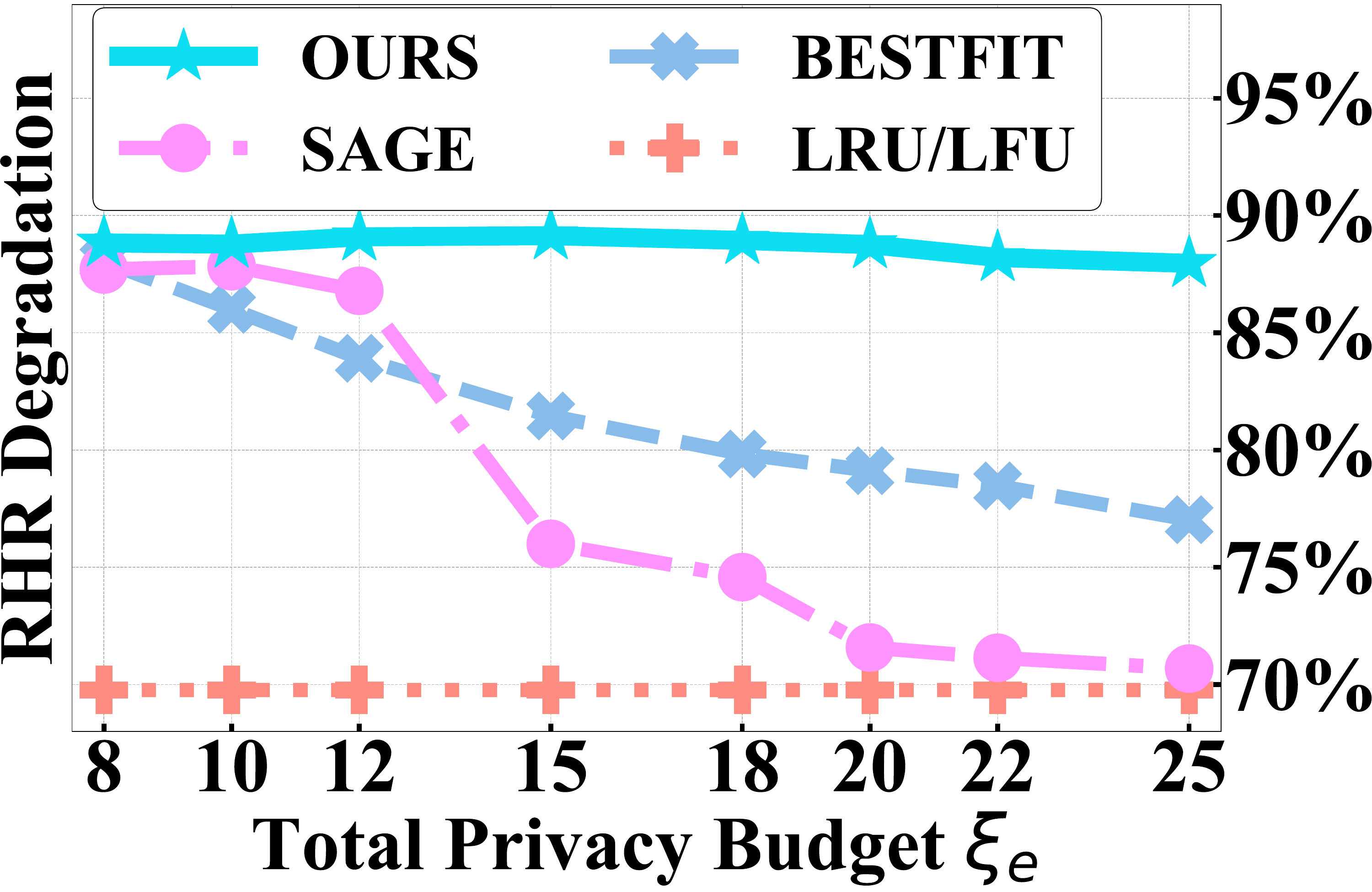}
    \caption{The average decline in RHR between users' origin and distorted profile when varying total privacy budget $\xi_e$ at EDs.} 
    \label{fig:xi_change_DRHR}
\end{subfigure}
\end{minipage}

\caption{Comparing privacy protection performances under different metrics for PPVF and other baselines with different settings of pre-fetching capacity or total privacy budget. A lower PDR and a larger RHR degradation are preferred by users to protect their private viewing profiles.}
\vspace{-4mm}
\end{figure*}

\subsection{Effectiveness of Privacy Protection} 
We first evaluate the performance of privacy protection 
using 
two metrics, 
i.e., the average JS and the degradation of RHR. 
We then 
further investigate the final status of the remaining privacy budget of all content.

\subsubsection{\textbf{Effectiveness of Distorting Private Information}}\label{SEC: Privacy Protection RESULTS JS}

For experiments presented in 
Fig.~\ref{fig:fe_change_jaccard} and Fig.~\ref{fig:xi_change_jaccard}, we compare the average JS between users' original profiles and profiles exposed by EDs after the test period.  
In Fig.~\ref{fig:fe_change_jaccard}, we fix \(c_e=1\%\) and \(\xi_e=15\), but vary \(f_e\) to study the average JS under different numbers of redundant requests. Whereas, in Fig.~\ref{fig:xi_change_jaccard}, we vary  \(\xi_e\) but fix \(c_e=1\%\) and \(f_e=4\) to study privacy protection under different privacy budget of EDs. 
As presented in Fig.~\ref{fig:fe_change_jaccard} and Fig.~\ref{fig:xi_change_jaccard}, we can observe that PPVF steadily outperforms other baselines. It exhibits an average reduction of $17.54\%$ ($22.38\%$) of JS compared to the second-best privacy-preserving baseline when varying the pre-fetching capacity (or the total privacy budget) for each ED. 

These experimental results manifest that PPVF can 
significantly distort exposed user profiles so that users' video request privacy can be preserved. 
Compared with SAGE and BESTFIT, PPVF 
achieves the lowest average JS because PPVF considers both limited privacy budget and video utility when pre-fetching videos. 
Recall that we tune a threshold to select videos for generating redundant requests in Alg.~\ref{alg: online allocation algorithm}. When the privacy budget is plentiful, PPVF selects videos of high utility with higher priority. However, if the privacy budget of the video is tight, we tune the threshold so that PPVF can select the video more conservatively. Instead, more diversified videos will be selected to conceal user privacy.  
Note that the average JS of classical caching algorithms, i.e.,  LRU, and LFU, are also compared with ours. Although these algorithms do not consider privacy protection with the redundant fetching videos, they can benchmark the degree of protection offered by privacy-preserving algorithms at edge devices.

\subsubsection{\textbf{Privacy Protection against Recommendation Systems}}\label{SEC: Privacy Protection RESULTS RHR}

We further employ the degradation of RHR to evaluate privacy protection by implementing the algorithm in~\cite{He2017} to recommend videos based on request records exposed by EDs after the test period. 
The configurations in Fig.~\ref{fig:fe_change_DRHR} and Fig.~\ref{fig:xi_change_DRHR} mirror those in Fig.~\ref{fig:fe_change_jaccard} and Fig.~\ref{fig:xi_change_jaccard}, respectively. 
By using original user profiles for a recommendation, the algorithm in~\cite{He2017} can achieve $99.42\%$ RHR, indicating the effectiveness of the recommendation. Then, the Degradation of RHR calculates the gap between the accuracy achieved by utilizing request records exposed by EDs and the original accuracy $99.42\%$. 
The experimental results in Fig.~\ref{fig:fe_change_DRHR} and Fig.~\ref{fig:xi_change_DRHR} indicate that:
\begin{itemize}
\item  PPVF is the best one 
to 
achieve the highest RHR degradation under all different scenarios. In particular, the performance of PPVF is better when the pre-fetching capacity is limited or the total privacy budget is sufficient.

\item SAGE and BESTFIT are better than LRU/LFU in privacy-preservation. However, their performance is inferior to PPVF under the same constraints, e.g., pre-fetching capacity and privacy budget. 

\item LRU/LFU can degrade RHR performance because they are implemented on EDs, which only expose consolidated request records of multiple users, making it difficult for recommenders to identify personalized interests. 
\end{itemize}

\begin{figure}[t]
\centering
\includegraphics[width=0.66\linewidth]{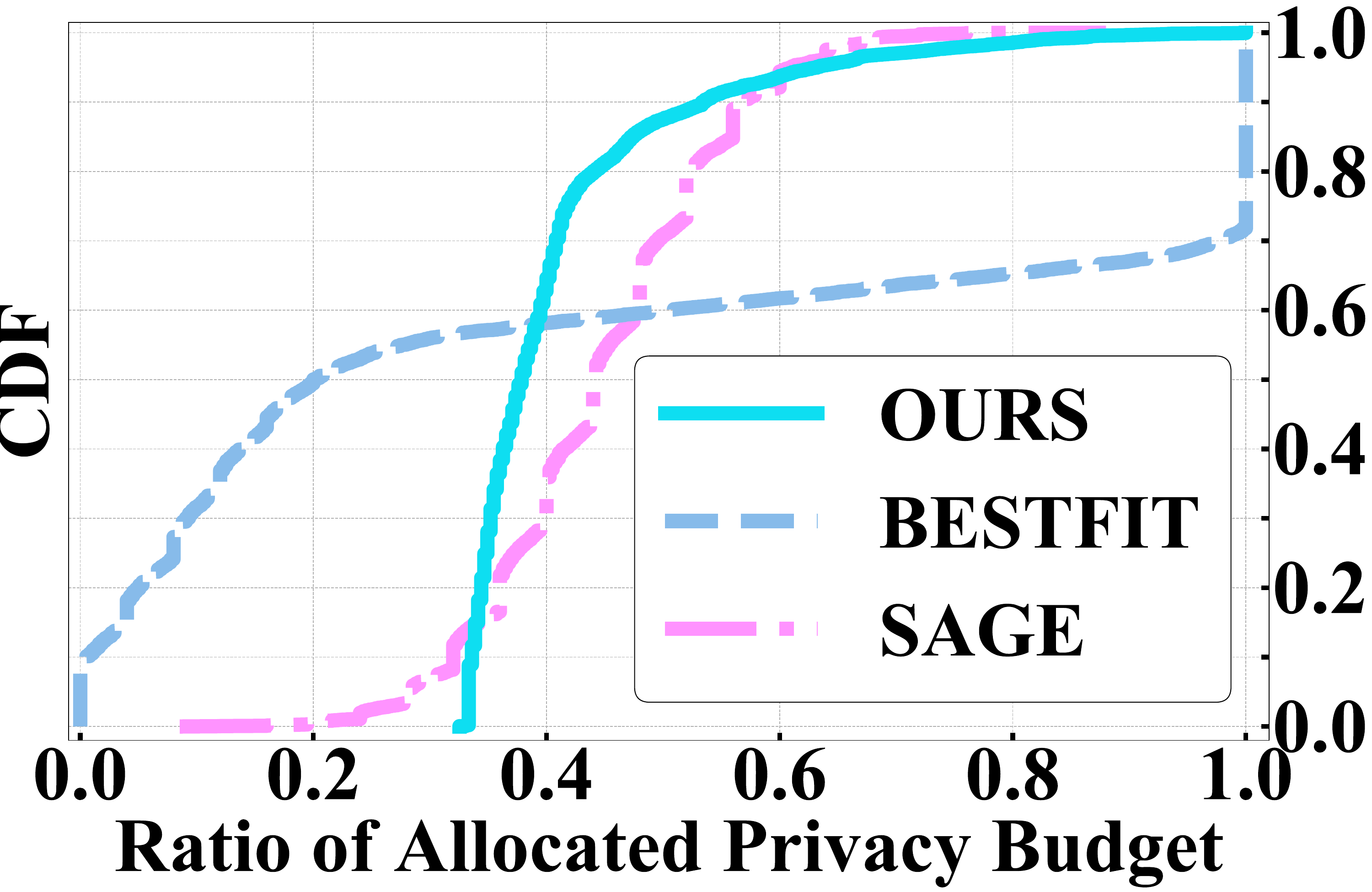}
    \caption{The CDF (cumulative distribution function) of the average ratio of the allocated privacy budget for videos at the end of the test.}
    \label{fig:remained_b}
\end{figure}

\begin{figure*}[!t]
\begin{subfigure}{0.47\linewidth}
\includegraphics[width=\linewidth]{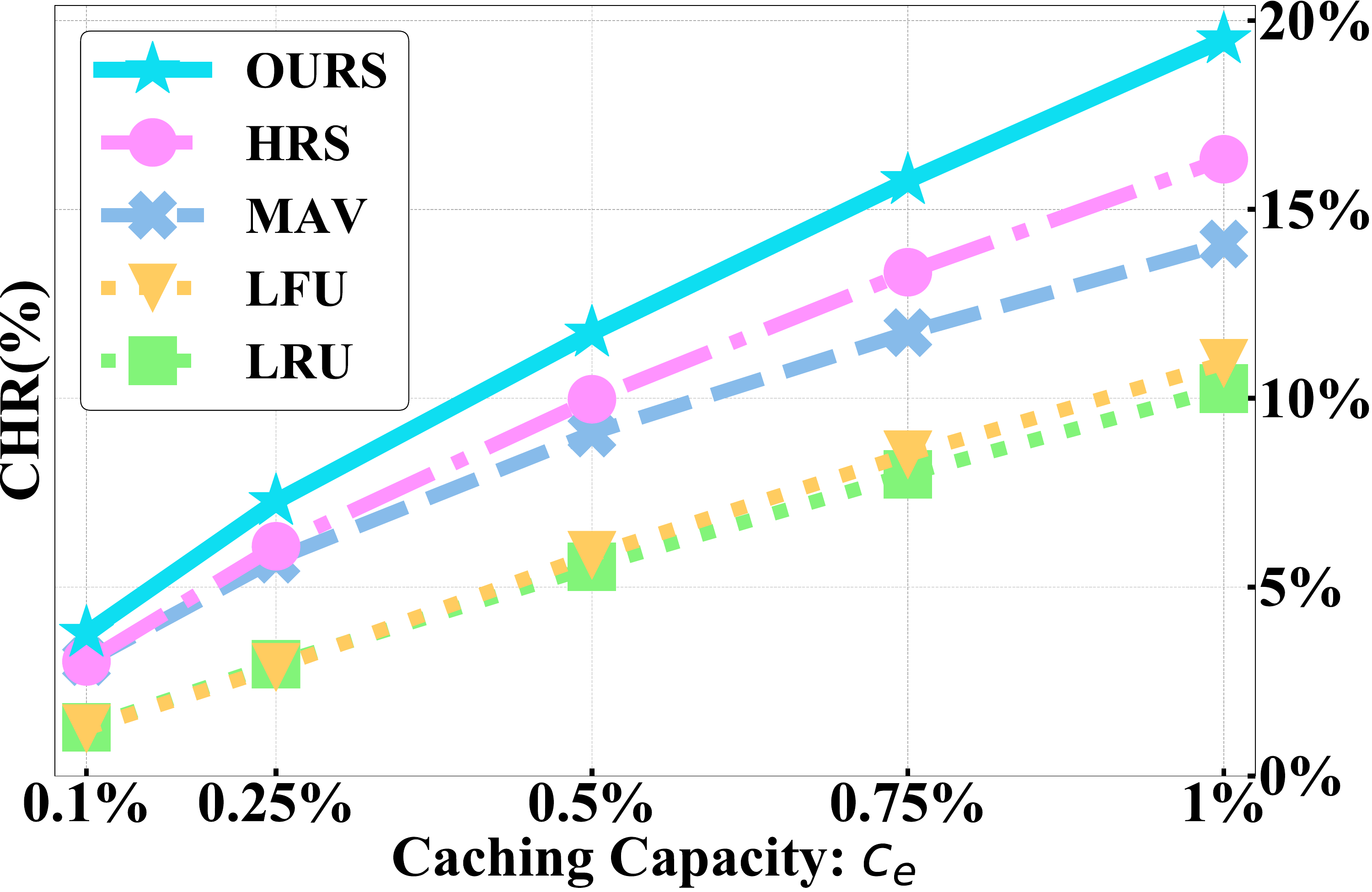}
    \caption{The average CHR  for all EDs in some small caching capacities. }
    \label{fig:hitrate_ce_1}
\end{subfigure}
\hfill
\begin{subfigure}{0.47\linewidth}
\includegraphics[width=\linewidth]{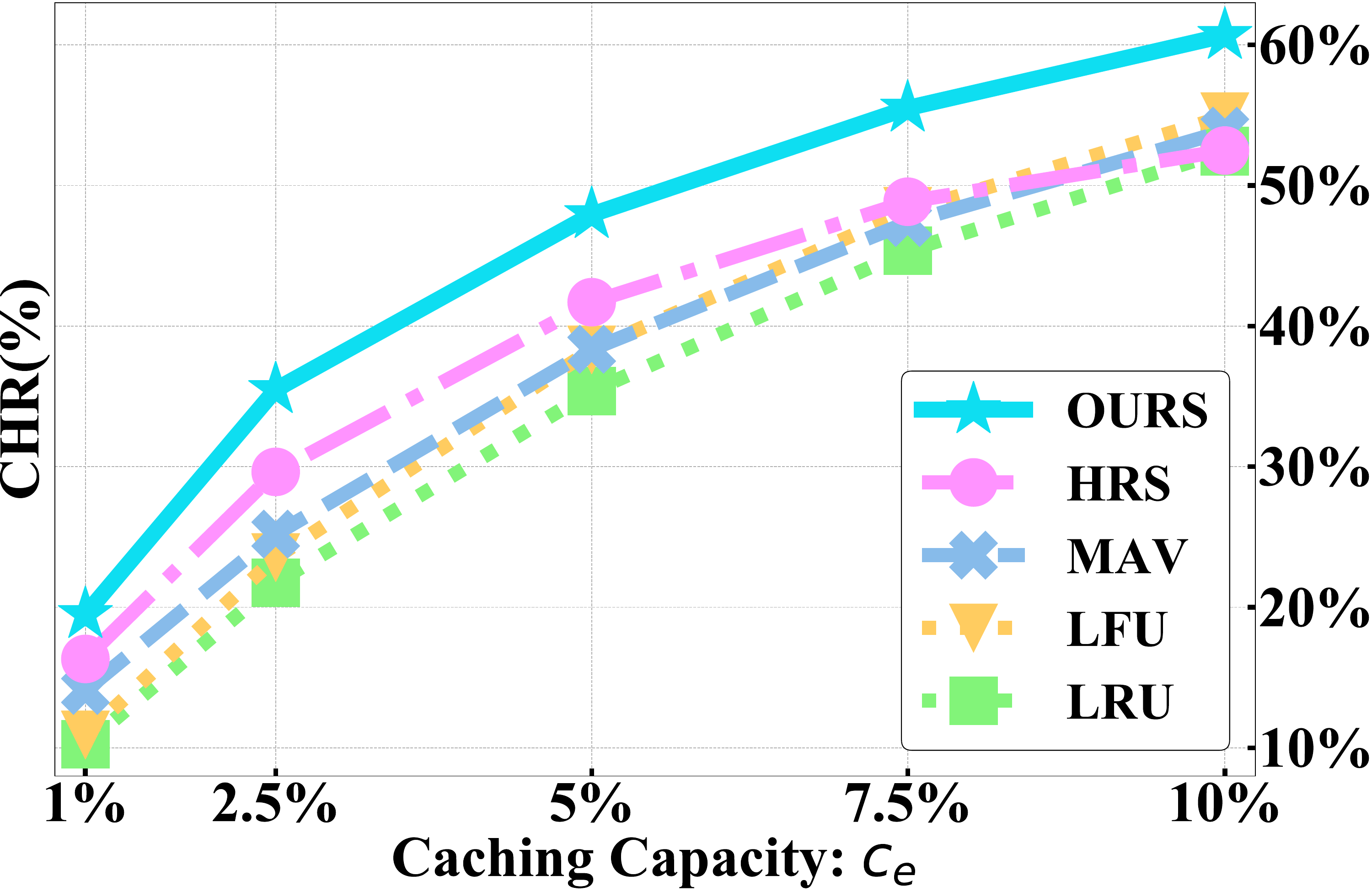}
\caption{The average CHR for all EDs in some large caching capacities. }
\label{fig:hitrate_ce}
\end{subfigure}
\caption{Comparing caching performance for PPVF and other baselines when varying the 
caching capacity $c_e$ of all EDs.  For more detailed results, please refer to Table~\ref{table: ce CHR} in Appendix~\ref{appendix: CHR}.}
\label{fig:hitrate}
\end{figure*}

\subsubsection{\textbf{Remaining Privacy Budgets of all Content}}
In Fig.~\ref{fig:remained_b}, we plot the cumulative distribution function (CDF) of the remaining privacy budgets of all videos after the test period to visualize redundant request decisions made by different caching algorithms. 
Here, we set $\xi_e=15$ by default. From  Fig.~\ref{fig:remained_b}, we can observe that PPVF can use up privacy budgets of videos worth for caching, and thereby, there are nearly $70\%$ videos with $60\%$ residual privacy budget at the end of the test. In contrast, the budget consumption of  BESTFIT and SAGE is scattered among different videos.  Their redundant requests for the hottest or coldest videos are 
not effective 
in preserving privacy, 
which is 
why their protection is weaker than ours.  


\subsection{Effective Caching Performance}\label{sec: caching results}

To comprehensively demonstrate the superiority of PPVF, we also compare CHR performance between different algorithms. 
In this experiment, we tune the caching capacity of each ED from $0.1\%$ to $10\%$ of the total video number, i.e.,  the caching capacity ranges from $10$ to $1037$ videos. For simplicity, we ignore the size difference of videos~\cite{Zhang2022}. 
We calculate the CHR of the entire system over the test period. The results are plotted in 
Fig.~\ref{fig:hitrate}, in which the y-axis represents the average CHR and the x-axis represents the caching capacity of each ED. 
To better show the difference between PPVF,  SAGE, and BESTFIT, we present numerical results in Table~\ref{table: ce CHR} in Appendix~\ref{appendix: CHR}. 
From the experimental results in Fig.~\ref{fig:hitrate} and Table~\ref{table: ce CHR}, we can observe that:
\begin{itemize}
\item PPVF is slightly better than SAGE and BESTFIT in terms of the CHR performance among the small capacities, while BESTFIT achieves the highest CHR when the caching capacity is large. Note that it is fair to compare the CHR performance between PPVF, SAGE, and BESTFIT because all algorithms cache redundant video requests based on the same predicted utility. 

\item  Except for SAGE and BESTFIT, our PPVF consistently attains the highest CHR in comparison to other caching baselines. 
This translates to an average enhancement of $18.15\%$ over the second-best caching algorithm, 
HRS, within all capacity settings. 
The presented results demonstrate the robustness of PPVF,  indicating its potential applicability for implementation across heterogeneous edge devices with varying caching capacities.

\item On the one hand, PPVF outperforms HRS / MAV in CHR by leveraging a more effective utility prediction method to reliably aggregate the private information at all EDs based on the FL framework. On the other hand, the CHR performance of PPVF surpasses that of LFU/LRU because these eviction-based algorithms do not make any redundant video requests to improve their caching efficiency.  The superior performance of PPVF over these SOTA baselines can be attributed to its ability to request and cache high-utility redundant videos, thereby elevating the CHR.

\item Moreover, PPVF demonstrates a more efficient utilization of caching capacity. To elucidate,
when working with a limited caching capacity of $0.1\%,$ PPVF's performance surpasses the second-best caching solution by more than $24.70\%$. This is especially notable under constrained caching resources, underscoring PPVF's exceptional capability to predict the most popular videos, even if using distorted information in the online FL framework. 
With more accurately estimated video utility, PPVF can accordingly pre-fetch videos to attain superior CHR performance.  
\end{itemize}

Lastly, we study the sensitivity of the online parameter update interval $t_\theta$ in Table~\ref{tab: t_theta} to see how this hyper-parameter affects the video caching performance. All other hyper-parameters are kept unchanged as we vary $t_\theta$.
As illustrated in Table~\ref{tab: t_theta}, the comprehensive CHR exhibits an effective improvement when this hyper-parameter is minimized. A smaller $t_\theta$ means a more frequent online parameter update in the FL framework. This observation stems from the fact that a more frequent update contributes to sustaining an efficient model for predicting video utility. Nevertheless, this enhancement comes at the expense of an increased computational burden. Considering this trade-off, the selection of a 2-day (48-hour) interval for parameter updates in previous work~\cite{Zhang2022} and our study is deemed rational.

\begin{table}[!tbp]
\renewcommand{\arraystretch}{1.5}
\setlength\tabcolsep{8pt}
\caption{The average CHR (\%) results under different $t_\theta$ (hours).  All Settings are on default as described in Sec.~\ref{SEC:Dataset and Settings} except varying except $t_\theta$.}
\begin{center}
\begin{tabular}{|p{20pt}<{\centering}||c|c|c|c|c|c|}
\hline
\multirow{2}{*}{\textbf{$c_e$}}&\multicolumn{6}{c|}{\normalsize{$t_\theta$ (hours)}} \\
\cline{2-7} 
&\makecell[c]{12}&\makecell[c]{24}&\makecell[c]{48}&\makecell[c]{72}&\makecell[c]{120}&\makecell[c]{240}\\
\hline
0.1\%&3.944&\textbf{3.975}&3.782&3.471&3.251&2.472\\
\hline
1\%&19.01&19.40&\textbf{19.49}&19.37&18.74&17.48\\
\hline
10\%&58.07&60.37&\textbf{60.68}&60.36&60.53&59.06\\
\hline
\end{tabular}
\label{tab: t_theta}
\end{center}
\end{table}

\section{RELATED WORK}\label{Sec:Related}
User privacy, including historical records, location, and other personal information, 
is 
an important concern in online video systems, prompting significant research efforts for safeguarding it. Various approaches have been introduced to address this concern. For instance, noise-based methods like DP~\cite{Zhang2022a} and Anonymous~\cite{Nisha2022, Hu2018} have been proposed to shield location information, while blockchain-based techniques~\cite{Dai2020, Cui2022} have been employed to safeguard users' personal information. Despite these efforts, the focus on request privacy, deemed the most critical user privacy aspect~\cite{Wang2019}, is also essential during the design of privacy-preserving video systems. In this section, we briefly review existing relevant works from two perspectives: privacy leakage in online video services and its protection.

\subsection{Request Privacy Leakage}

Privacy concerns in online video services span multiple dimensions: request traces~\cite{Tong2022, Sivaraman2021}, personal details~\cite{Xue2019, Zhang2022b}, location data~\cite{Hu2018, Amini2011, Nisha2022}, and specific content data~\cite{Araldo2018, Cui2020}, to name a few. Among these, request traces have emerged as particularly pivotal within online video services, as they may inadvertently expose user preferences to potential adversaries~\cite{Wang2019}. Such traces frequently encapsulate sensitive user information, capturing browsing patterns, preferences, and interests of users~\cite{Sivaraman2021, Cui2020}.
Commercial motivations propel content providers to amass and scrutinize users' private data~\cite{Ni2021}. This collected data is multifaceted, comprising geographical locations~\cite{Amini2011}, behavioral tendencies~\cite{Zhang2019a}, personal specifics~\cite{Zhou2019}, among other aspects. Leveraging this data can substantially refine service quality for content providers across domains, including content caching~\cite{Shi2021}, recommendation engines~\cite{Guerraoui2017}, and video distribution~\cite{Gupta2022, Kirilin2020}.

\subsection{Request Privacy Protection at the Network Edge}
Privacy protection in online video services at the network edge can be broadly classified into three categories. The \emph{first} focuses on cryptography based techniques, including encryption transmission~\cite{Araldo2018, Xu2019} and blockchain-enabled methods~\cite{ Qian2020, Jiang2022}. While these techniques are potent, they impose significant computational demands on edge devices and cannot entirely prevent CPs from potential misuse of user data.
The \emph{second} category encompasses trusted distributed computing (TDC) techniques, exemplified by federated learning~\cite{Qiao2022, Liu2023,Liu2022}. Although these methods bolster user privacy by obviating the need for direct data transfer, their suitability for online video platforms is debatable, given their limited capability to prohibit content providers from tracking user viewing habits.
The \emph{third} category is grounded in noise-based techniques. These methods accentuate request privacy within edge networks by obfuscating actual user preferences~\cite{Sen2018, Wang2022a}. A prevalent approach within this category is the pre-fetching of redundant and unrelated videos to foster ambiguity. Such pre-fetched content can also be cached at the network edge to serve future requests, thereby curtailing direct interactions with CPs~\cite{Wu2016, Nikolaou2016}, which in turn mitigates the data exposure risk. Nevertheless, balancing the quality of edge services with the imperative of user data protection remains an intricate endeavor.

one viable solution to ensure request privacy in online video systems involves incorporating DP noises, which delivers robust information protection assurances~\cite{Cai2023, Wang2019, Zhou2019}. L{'e}cuyer et al.~\cite{Lecuyer2019} pioneered the use of block composition to tackle privacy concerns arising from expanding private datasets. This innovative method provides theoretical assurances for the efficient utilization of individual dataset segments. Moreover, to shield non-iid datasets, correlated differential privacy was introduced in~\cite{Chen2017, Zhu2015}, taking into account the interdependence among records. Yet, the challenge confronted by allocating limited privacy budgets for online video requests on edge devices persists. Certain allocation frameworks, like Sage~\cite{Lecuyer2019} and DP-FLames~\cite{Hu2023}, may be overly simplistic or rely on improbable assumptions, thereby restricting their flexibility in diverse scenarios.
In light of these methodological limitations, we present a novel privacy protection strategy. This method enhances request privacy by generating redundant requests, all while preserving the operational efficacy of edge caching.

\section{Conclusion}\label{sec: conclusion}
With the proliferation of online video services, preserving request privacy remains an open problem. 
The challenge of this problem lies in that online video providers can automatically capture video requests from users.  As a consequence, user requests cannot be trivially distorted by injecting noises or protected with encryption. 
In this work, we are among the first to attempt to address this challenge by proposing the PPVF framework, which synthetically utilizes trusted edge caching, correlated differential privacy, and federated learning. 
In other words, edge devices try to conceal user request privacy by generating noisy requests (with the noise scale calibrated according to video correlation) to the video provider. To maintain system efficiency, edge devices collaboratively predict video utility via FL so that 
they 
can harmonize video utility and privacy leakage amount when requesting videos.  
With the advancement of the online video market, privacy-preserving techniques presented in this work offer invaluable insights and solutions for bolstering user privacy when consuming video content. 
Subsequent endeavors can build upon these foundations to further propel the field of privacy-preserving online video services.




%% file: bio.tex
\begin{IEEEbiography}
[{\includegraphics[width=1in,height=1.25in,clip,keepaspectratio]{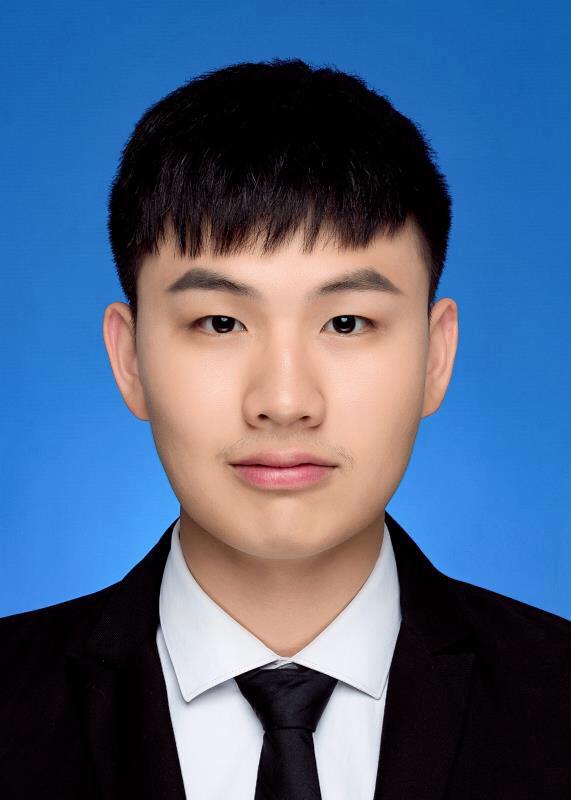}}]
{Xianzhi Zhang} received his B.S. degree from Nanchang University (NCU), Nanchang, China, in 2019 and an M.S. degree from the School of Computer Science and Engineering, Sun Yat-sen University (SYSU), Guangzhou, China, in 2022.  He is currently pursuing a Ph.D. degree in the School of Computer Science and Engineering at Sun Yat-sen University, Guangzhou, China. He is also working as a visiting PhD student at the School of Computing,  Macquarie University, Sydney, Australia. Xianzhi's current research interests include video caching, caching privacy, machine learning, and edge computing. His research has been published at IEEE TPDS and won the Best Paper Award at PDCAT 2021. 
\end{IEEEbiography}

\begin{IEEEbiography}[{\includegraphics[width=1in,height=1.25in,clip,keepaspectratio]{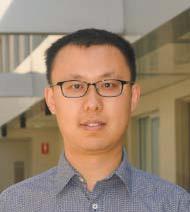}}]
{Yipeng Zhou} is a senior lecturer in computer science at the School of Computing, Macquarie University, and the recipient of ARC (Australian Research Council) DECRA in 2018. From Aug. 2016 to Feb. 2018, he was a research fellow with the Institute for Telecommunications Research (ITR) of the University of South Australia. From 2013.9-to 2016.9, He was a lecturer at the College of Computer Science and Software Engineering at Shenzhen University. He was a Postdoctoral Fellow with the Institute of Network Coding (INC) of the Chinese University of Hong Kong (CUHK) from Aug. 2012 to Aug. 2013. He obtained his PhD degree and MPhil degree from the Information Engineering (IE) Department of CUHK, respectively. He got a Bachelor's degree in Computer Science from the University of Science and Technology of China (USTC). His research interests include federated learning, privacy protection, and caching algorithm design in networks. He has published more than 80 papers including IEEE INFOCOM, ICNP, IWQoS, IEEE ToN, JSAC, TPDS, TMC, TMM, etc.
\end{IEEEbiography}

\begin{IEEEbiography}[{\includegraphics[width=1in,height=1.25in,clip,keepaspectratio]{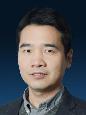}}]
{Di Wu}(M'06-SM'17) received his B.S. degree from the University of Science and Technology of China, Hefei, China, in 2000, the M.S. degree from the Institute of Computing Technology, Chinese Academy of Sciences, Beijing, China, in 2003, and the Ph.D. degree in computer science and engineering from the Chinese University of Hong Kong, Hong Kong, in 2007. He was a Post-Doctoral Researcher with the Department of Computer Science and Engineering, Polytechnic Institute of New York University, Brooklyn, NY, USA, from 2007 to 2009, advised by Prof. K. W. Ross. Dr. Wu is currently a Professor and the Associate Dean of the School of Computer Science and Engineering at Sun Yat-sen University, Guangzhou, China. He was the recipient of the IEEE INFOCOM 2009 Best Paper Award and IEEE Jack Neubauer Memorial Award. His research interests include edge/cloud computing, multimedia communication, Internet measurement, and network security.
\end{IEEEbiography}

\begin{IEEEbiography}[{\includegraphics[width=1in,height=1.25in,clip,keepaspectratio]{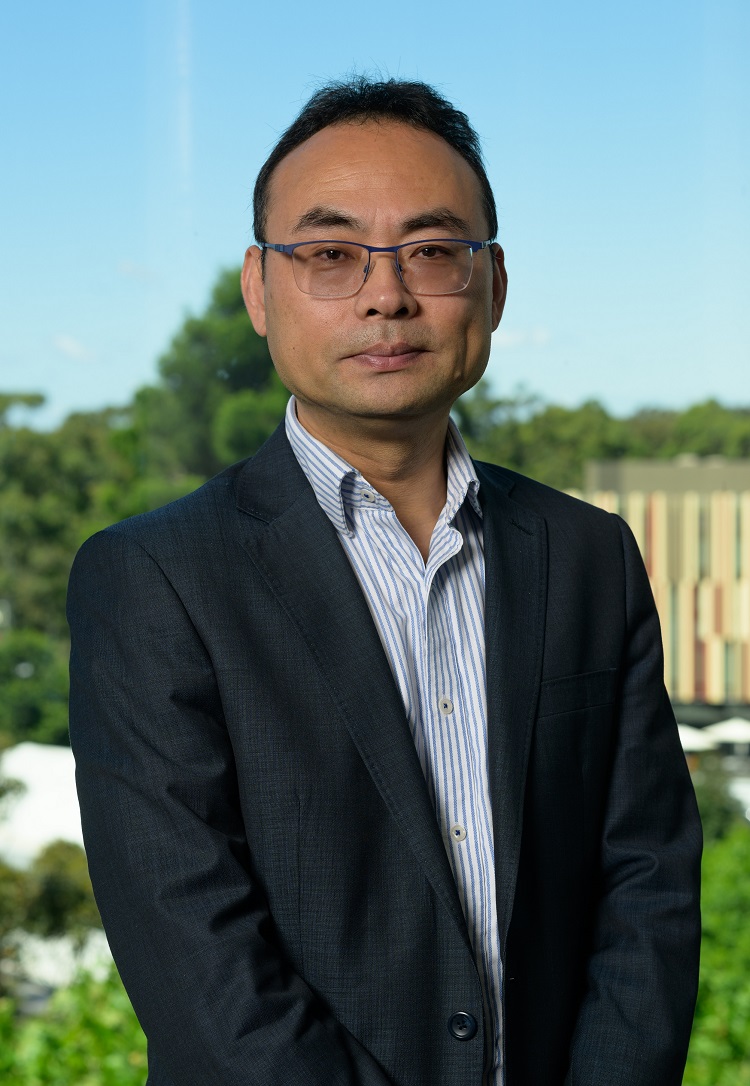}}]
{Quan Z. Sheng} is a Distinguished Professor and Head of the School of Computing at Macquarie University, Sydney, Australia. Before moving to Macquarie University, Michael spent 10 years at School of Computer Science, the University of Adelaide, serving in senior leadership roles such as Interim Head of School and Deputy Head of School. Michael holds a PhD degree in computer science from the University of New South Wales (UNSW) and did his post-doc as a research scientist at CSIRO ICT Centre. From 1999 to 2001, Michael also worked at UNSW as a visiting research fellow. Prior to that, he spent 6 years as a senior software engineer in industries. 

Prof. Michael Sheng's research interests include Web of Things, Internet of Things, Big Data Analytics, Web Science, Service-oriented Computing, Pervasive Computing, and Sensor Networks. He is ranked by Microsoft Academic as one of the Most Impactful Authors in Services Computing (ranked Top 5 all time worldwide) and in Web of Things (ranked Top 20 all time). He is the recipient of the AMiner Most Influential Scholar Award on IoT (2019), ARC (Australian Research Council) Future Fellowship (2014), Chris Wallace Award for Outstanding Research Contribution (2012), and Microsoft Research Fellowship (2003). Prof Michael Sheng is Vice Chair of the Executive Committee of the IEEE Technical Community on Services Computing (IEEE TCSVC), the Associate Director (Smart Technologies) of Macquarie's Smart Green Cities Research Centre, and a member of the ACS (Australian Computer Society) Technical Advisory Board on IoT.
\end{IEEEbiography}

\begin{IEEEbiography}[{\includegraphics[width=1in,height=1.25in,clip,keepaspectratio]{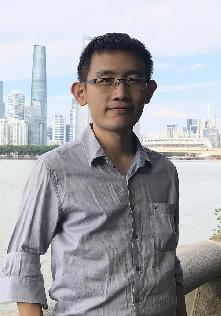}}]
{Miao Hu} (S'13-M'17) is currently an Associate Professor at the School of Computer Science and Engineering at Sun Yat-Sen University, Guangzhou, China. He received a B.S. degree and a Ph.D. degree in communication engineering from Beijing Jiaotong University, Beijing, China, in 2011 and 2017, respectively. From September 2014 to September 2015, he was a Visiting Scholar at the Pennsylvania State University, PA, USA. His research interests include edge/cloud computing, multimedia communication, and software-defined networks.
\end{IEEEbiography}

\begin{IEEEbiography}[{\includegraphics[width=1in,height=1.25in,clip,keepaspectratio]{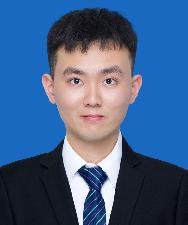}}]
{Linchang Xiao} received his B.S. Degree from the School of Computer Science and Engineering, Sun Yat-sen University (SYSU), Guangzhou, China, in 2022. He is currently working toward his M.S. degree at the School of Computer Science and Engineering, Sun Yat-sen University (SYSU), Guangzhou, China. His research interests include cloud \& edge computing, content caching, and multimedia communication. 
\end{IEEEbiography}

%% file: appendix.tex
\begin{appendices}


\section{Proof of Theorem \ref{theorem: CR}}\label{appendix: CR}
Before the proof, the detailed algorithm for online video selection and privacy budget allocation is shown in Alg.~\ref{alg: online allocation algorithm}.

\begin{proof}
We initiate our discussion with a single content scenario and subsequently extend our proof to a multi-video content context. Besides focusing on a particular ED, we simplify our notation by omitting the subscript \( e \), provided there is no risk of confusion.

For any given input sequence \( \kappa \), we define \( \text{PPVF}_i(\kappa) = \sum_{k\in \mathcal{S}_i} \lambda_i^k \) and \( \text{OPT}_i(\kappa) = \sum_{k\in \mathcal{S}^\star_i} \lambda_i^k \) as the total utilities accrued by the PPVF algorithm (as outlined in Alg.~\ref{alg: online allocation algorithm}) and the offline optimum (denoted as OPT), respectively. Here, \( \mathcal{S}_i \) and \( \mathcal{S}^\star_i \) represent the set of requests selected to video $i$ from the input sequence by these two methods.
Let \( \Gamma_i \) represent the fraction of video \( i \)'s budget consumed by PPVF. 

Furthermore, we define $$ \Lambda_i = \sum_{k\in(\mathcal{S} \cap \mathcal{S}^\star)} \lambda_{i}^k, \quad \Lambda'_i = \sum_{k\in(\mathcal{S} \backslash \mathcal{S}^\star)} \lambda_{i}^k,$$
and
\[ \Upsilon_i = \sum_{k\in(\mathcal{S} \cap \mathcal{S}^\star)} \Theta(\gamma_{i}^k)\cdot \epsilon_{i},\quad \Upsilon_i' = \sum_{k\in(\mathcal{S} \backslash \mathcal{S}^\star)}  \Theta(\gamma_{i}^k)\cdot \epsilon_{i}. \]

First, for each request \( k \in \mathcal{S}_i \), the efficiency \(\lambda_i^k/\epsilon_{i}\) is at least \( \Theta(\gamma_{i}^k) \), i.e., \( \lambda_i^k > \Theta(\gamma_{i}^k)\cdot \epsilon_{i} \), where \( \gamma_{i}^k \) denotes the fraction of privacy budget of video \( i \) accessed at that specific time.
Rounding down the utility \(\lambda_i^k\) of each request \( k \) chosen by PPVF to \( \Theta(\gamma_{i}^k)\cdot \epsilon_{i} \), we ascertain that 
\begin{subequations}\label{eq: th1 begin}
\begin{align}
    \Upsilon_i\leq\Lambda_i, \\ \Upsilon_i' \leq\Lambda'_i. 
\end{align}
\end{subequations}
Recall that \(\Theta(\gamma)\) is a monotonically increasing function with respect to \(\gamma\), we can also observe
\begin{equation}
\Upsilon_i \leq \Theta(\Gamma_i) \cdot E_i,
\end{equation}
where $ E_i = \sum_{k\in ( \mathcal{S} \cap \mathcal{S}^\star)}\epsilon_{i}$.

Continuing our analysis, for each request \( k \in \mathcal{S}_i^\star - (\mathcal{S}_i \cap \mathcal{S}_i^\star) \), which represents requests selected by OPT but not by our PPVF algorithm, we have:
\begin{equation}
    \lambda_{i}^k \leq \Theta(\gamma^k_{i}) \cdot \epsilon_{i} \leq \Theta(\Gamma_i) \cdot \epsilon_{i}.
\end{equation}
Note that \( \xi - E_i \) is the remaining budget as per PPVF after selecting requests to set \(\mathcal{S} \cap \mathcal{S}^\star\), which 
represents the ideal maximum budget that OPT could employ to select the request to the set \( \mathcal{S}^\star - \mathcal{S} \cap \mathcal{S}^\star \). Given the threshold function \(\Theta(\gamma)\) is 
monotonically increasing 
with respect to \(\gamma\), we can derive:
\begin{equation}
    \text{OPT}_i(\kappa) - \Lambda_i    \leq  \Theta(\Gamma_i) \cdot (\xi - E_i).
\end{equation}
Since $\text{PPVF}_i(\kappa)= \Lambda_i + \Lambda'_i$, the above inequality implies that 
\begin{equation}\label{eq: th1 end}
    \frac{\text{OPT}_i(\kappa)}{\text{PPVF}_i(\kappa)} \leq \frac{\Lambda_i + \Theta(\Gamma_i)\cdot (\xi_{e} - E_i)}{\Lambda_i + \Lambda'_i}.
\end{equation}

Additionally, considering $\text{OPT}_i(\kappa)\geq \text{PPVF}_i(\kappa)$, we always have $\Theta(\Gamma_i)\cdot (\xi - E_i)\geq\Lambda'_i$. Thus, if we reduce $\Lambda_i$ to $\Upsilon_i$ in both denominator and numerator of Eq.~\eqref{eq: th1 end}, the ratio of $\frac{\text{OPT}_i(\kappa)}{\text{PPVF}_i(\kappa)}$ increases. 
In conclusion, combining the inequations in Eqs.~\eqref{eq: th1 begin}-\eqref{eq: th1 end}, we have:
\begin{equation}\label{eq: OPT_i/ALG_i }
\begin{aligned}
    \frac{\text{OPT}_i(\kappa)}{\text{PPVF}_i(\kappa)} 
    &\leq \frac{\Upsilon_i + \Theta(\Gamma_i)\cdot (\xi - E_i)}{\Upsilon_i + \Lambda'_i} \\
    &\leq \frac{\Theta(\Gamma_i) \cdot E_i + \Theta(\Gamma_i)\cdot (\xi - E_i)}{\Upsilon_i + \Lambda'_i} \\
    &\leq \frac{ \Theta(\Gamma_i)\cdot \xi}{\Upsilon_i + \Upsilon_i'} 
    = \frac{ \Theta(\Gamma_i)}{\sum_{k\in\mathcal{S}} \Theta(\gamma_{i}^k)\Delta \gamma_{i}^k}.
\end{aligned}
\end{equation}
Recall that $\Gamma_e = \frac{1}{1+ln(U_e/L_e)}$ is the lowest threshold in function $\Theta_e(\cdot)$ at any ED $e$.
Based on 
Assumption~\ref{assumption: tiny cost}, indicating that $\Delta\gamma_{e,i}^k\rightarrow 0$, we have:
\begin{equation}
\begin{aligned}
&\sum_{k\in\mathcal{S}_{e,i}} \Theta_e(\gamma_{e,i}^k)\Delta \gamma_{e,i}^k \approx \int_{0}^{\Gamma_i} \Theta_e(\gamma)\ \mathrm{d} \gamma\\
= &\int_0^{\Gamma_e} L_e \cdot \mathrm{d} \gamma +\int_{\Gamma_e}^{\Gamma_i} \frac{L_e}{\exp(1)} \ln\left(\frac{U_e \cdot \exp(1)}{L_e}\right)^{\gamma} \cdot \mathrm{d} \gamma \\
 =&\Gamma_e \cdot \left( L_e  + \frac{L_e}{\exp(1)} \left(\ln\left(\frac{U_e \cdot \exp(1)}{L_e}\right)^{\Gamma_i}\right.\right.\\
 &\hspace{100pt} -\left.\left.\ln\left(\frac{U_e \cdot \exp(1)}{L_e}\right)^{\Gamma} \right)\right)\\
 =&\Gamma_e \cdot \frac{L_e}{\exp(1)} \ln\left(\frac{U_e \cdot \exp(1)}{L_e}\right)^{\Gamma_i}\\
 =&\Gamma_e \cdot \Theta_e(\Gamma_i).
\end{aligned}
\end{equation}

Therefore, the ratio $\frac{\text{OPT}_i(\kappa)}{\text{PPVF}_i(\kappa)}$ at ED $e$ can be derived into
\begin{equation}\label{eq: eventual OPT_i/ALG_i}
\begin{aligned}
    \frac{\text{OPT}_i(\kappa)}{\text{PPVF}_i(\kappa)} 
    &\leq \frac{ \Theta(\Gamma_i)}{\Gamma_e \cdot \Theta(\Gamma_i)}
    = 1+\ln(U_e/L_e).
\end{aligned}
\end{equation}
Denote $\mathcal{J}_{e}^{*}$ and $\mathcal{J}_{e}^{\text{PPVF}}$ as the final sum of utility obtained by solution of offline optimum and PPVF, respectively.
Following the proof in~\cite{Li2023,Li2021}, the CR of our online algorithm at any ED $e$ can be obtained by by summing up
all single video $i$:
\begin{equation}\label{eq: eventual OPT/ALG }
\begin{aligned}
\mathcal{J}_{e}^{*} &= \sum_i \text{OPT}_i(\kappa) \\
&\leq \sum_i \left(1+\ln(U_e/L_e)\right) \cdot  \text{PPVF}_i(\kappa) \\
&=\left(1+\ln(U_e/L_e)\right) \cdot\mathcal{J}_{e}^{\text{PPVF}}.
\end{aligned}
\end{equation}

To sum up, we can similarly prove that Alg.~\ref{alg: online allocation algorithm} achieves the best competition 
ratio
(CR) among all online solutions under Assumption~\ref{assumption: tiny cost}. 
\begin{equation}\label{eq: CR}
    CR=\max_{\kappa}\frac{\mathcal{J}_{e}^{*}}{\mathcal{J}_{e}^{\text{PPVF}}} = \left(1+\ln(U_e/L_e)\right).
\end{equation}
Proof completes.
\end{proof}


\begin{table*}[!tb]
\renewcommand{\arraystretch}{1.5}
\setlength\tabcolsep{3pt}
\caption{CHR (\%) under different caching capacity $c_e$. SAGE and BESTFIT are implemented with our video utility predictor. }
\resizebox{0.97\linewidth}{!}{
\begin{tabular}{
p{2cm}<{\centering}||
p{1.5cm}<{\centering}|p{1.5cm}<{\centering}|
p{1.5cm}<{\centering}|p{1.5cm}<{\centering}|
p{1.5cm}<{\centering}|p{1.5cm}<{\centering}|
p{1.5cm}<{\centering}|p{1.5cm}<{\centering}|
p{1.5cm}<{\centering}}
\toprule
\cline{1-10} 
\textbf{Caching Capacity}&\multirow{2}{*}{\textbf{0.1\%}} &\multirow{2}{*}{\textbf{0.25\%}}&\multirow{2}{*}{\textbf{0.5\%}}&\multirow{2}{*}{\textbf{0.75\%}}&\multirow{2}{*}{\textbf{1\%}}&\multirow{2}{*}{\textbf{2.5\%}}&\multirow{2}{*}{\textbf{5\%}}&\multirow{2}{*}{\textbf{7.5\%}}&\multirow{2}{*}{\textbf{10\%}}\\
\midrule
\cline{1-10} 
PPVF&\textbf{3.782} &\textbf{7.270} & \textbf{11.728}  &\textbf{15.764} &\textbf{19.492} &35.487  &47.936 & 55.505 & 60.677  \\
SAGE&3.768&7.266 &11.719 &15.761 &19.491 &35.431 &47.905 &55.451 &60.569 \\
BESTFIT&3.779 &7.256 &11.710 &15.688 &19.431 &\textbf{35.762 }&\textbf{48.394} & \textbf{55.971} & \textbf{61.528}  \\
HRS & 3.033&6.082&9.974&13.336&16.327&29.632&41.700&48.845&52.483 \\
MAV & 3.029&5.740&9.079&11.737&14.083&25.209&38.351&47.365&53.862 \\
LRU&1.293&2.965&5.539&7.989&10.257&21.739&35.366&45.367&52.436 \\
LFU&1.256&2.878&5.847&8.506&10.925&23.542&38.312&48.171&54.896 \\
\cline{1-10} 
\bottomrule
\end{tabular}
\label{table: ce CHR}
}
\vspace{-3mm}
\end{table*}

\begin{algorithm}[!t]
\SetKwFunction{localtraining}{EDLocalTraining}
\SetKwProg{Fn}{Function}{:}{}
\caption{Online and distributed parameters learning algorithm for MEP.}\label{Online Parameters Learning algorithm}
\label{alg:Online train}
\KwIn{ Online update time set $\mathcal{Q}$, Update interval $\Delta t$}
\KwOut{$\bm{\theta}$}
\For{$t_{\theta} \in \mathcal{Q} $}{
$n \leftarrow 0$ and initialize $\bm{\theta}^{(n)}$ with the outdated parameters;\\
\While{The termination condition is not satisfied}{
    $l_G \leftarrow 0$, $\bm{\nabla}_G \leftarrow [0]^{(2\times\text{D}+1)\times \text{I}}$;\\
    \For{$\forall e \in \mathcal{E} $}{
        $l_e$, $\bm{\nabla}_e \leftarrow$\localtraining{ $t_{\theta}$, $\Delta t$, $\bm{\theta}^{(n)}$};\\
        $l_G \leftarrow l_G + l_e$,$\bm{\nabla}_G \leftarrow \bm{\nabla}_G + \bm{\nabla}_e$;\\
    }
    Update $\bm{\theta}^{(n+1)}\leftarrow \bm{\theta}^{(n)}$ with $l_G$, $\bm{\nabla}_G$ and the penalty term $\bm{\rho}$; \\
    $ n \leftarrow n + 1$;  \\
    }
}
\Return $\bm{\theta}$

\Fn{\localtraining{ $t_{\theta}$, $\Delta t$, $\bm{\theta}^{(n)}$\label{fun:ED local training}}}{

$l_e\leftarrow 0$,
$\bm{\nabla}_{\beta} \leftarrow [0]^{\text{I}}$,
$\bm{\nabla}_{p} \leftarrow [0]^{\text{I} \times \text{D}}$,
$\bm{\nabla}_{q} \leftarrow [0]^{\text{I} \times \text{D}}$;\\

\For{$\forall i\in \mathcal{I} $}{
Collect the timestamp set of historical viewing requests $\mathcal{T}^{t_{\theta}-\Delta t, t_{\theta}}_{e,i}$ from the local dataset;\\
    \For{$\forall \tau \in \mathcal{T}^{t_{\theta} -\Delta t,t_{\theta}  }_{e,i} $
    }{
        Obtain the intensity $\hat{h}_{e}(i,\tau)$ by Eq.~\eqref{EQ: lambda_{e,i}^k com};\\
            $l_e \ \leftarrow l_e + \log \hat{h}_{e}(i,\tau)$;\\
        } 
    Calculate the term of integration $l'= \int_{t_{\theta}-\Delta t}^{t_{\theta}} \hat{h}_e(i,t) \mathrm{d} t$ in Eq.\eqref{EQ:logL online};\\
    $l_e \leftarrow l_e - l'$;\\
    Calculate the gradient $\nabla_\beta[i]$, $\bm{\nabla}_{p}[i]$ by Eqs.~\eqref{eq: ll dev p i} and \eqref{eq: ll dev betai}, respectively;\\
    \For{$\forall j\in \mathcal{I} $}{
        Calculate the gradient $\bm{\nabla}_{q}[j]$ by Eq.~\eqref{eq: ll dev q i};\\
        }
    }
Combine $\bm{\nabla}_\beta$, $\bm{\nabla}_{p}[i]$, $\bm{\nabla}_{q}[j]$ to gain the whole $\bm{\nabla}_e$;\\
\Return $l_e$, $\bm{\nabla}_e$
}
\end{algorithm}
\vspace{-3mm}

\vspace{-4mm}
\section{Detailed Derivation for log-likelihood function}~\label{appendix:log-likelihood}
Let the occurrence time $t^{\nu}$ be the time of the last viewing request in the historical viewing request set $\mathcal{V}_{e}^\nu$.
Given the overall intensity function $\hat{h}'_e(t) = \sum_{\forall i}\hat{h}_{e}(t,i)$ for any ED $e$, the probability that no request occurs in the period $[\,t^{\nu}, t\,)\ ,\ t<t^{\nu+1}$  is 
$$P\left\{\text{no request occurs in } [\,t^{\nu}, t\,)\ \big| \ \mathcal{V}_{e}\right\} = \exp \left[-\int_{t^\nu}^{t} \hat{h}'_e(t') \mathrm{d} t'\right].$$

Thus, the probability that a request for video $i$ occurs at time $t$ is given by 
$$P\left\{i,t \ \big| \ \mathcal{V}_e^{\nu}\,\right\}=\hat{h}_{e}(t,i) \exp \left[-\int_{t^{\nu}}^{t} \hat{h}'_e(t') \mathrm{d} t'\right].$$

For convenience, we let $t^0 = 0$ and align the all-time series for different videos $i$ to the same initial point. 
Recall that $\mathcal{V}_e$ as the historical dataset of all viewing requests at ED $e$ between the time $(\,0, T\,]$, it is easy to derive the likelihood function for all the parameters $\bm{\theta}\defeq\{\bm{p},\bm{q},\bm{\beta}\ \}$ shown in Eq.~(\ref{Eq: L}).
\begin{equation}
    \begin{split}
        l_e \left( \, \bm{\theta} \, \big| \, \mathcal{V}_e \, \right)
        &= \prod_{\nu=1}^{|\mathcal{V}_e|}\hat{h}_{e}(i,t^\nu)
          \exp \left[-\int_{t^{\nu-1}}^{t^{\nu}} \hat{h}'_{e}(t) \mathrm{d}t \right]\\ 
          &\cdot \exp \left[ -\int_{t^{|\mathcal{V}_e|}}^{T} \hat{h}'_{e}(t) \mathrm{d}t\right] ,\\
        &= \prod_{i\in \mathcal{I}}\prod_{\tau \in \mathcal{T}_{e,i}}  \hat{h}_{e}(i,\tau) \cdot \exp \left[ -\int_{0}^{T} \hat{h}_{e}(t) \mathrm{d} t\right] .
    \label{Eq: L}
\end{split}
\end{equation}
where $\mathcal{T}_{e,i}$ denoted the timestamp set of the viewing requests to video $i$ arriving at ED $e$ in $[0,T)$. In order to facilitate calculation, we can further derive the log-likelihood function to optimize all the parameters $\bm{\theta}$, which is defined as:
\begin{equation}
\begin{split}
    ll_e \left( \, \bm{\theta} \, \big| \, \mathcal{V}_e \, \right) =   \sum_{i\in \mathcal{I}} \sum_{ \tau \in \mathcal{T}_{e,i}} \log \hat{h}_{e}(i,\tau)(\tau) - \int_{0}^{T} \hat{h}_{e}(i,t') \mathrm{d} t'. 
\end{split}
\end{equation}


\section{FL-based online parameters estimation algorithm}\label{appendix: partial derivatives}

The detailed algorithm for online FL-based parameter estimation for the MEP model is presented in Alg.~\ref{alg:Online train}.

\section{Supplement of CHR Results}\label{appendix: CHR}
To better display the caching performance difference among different baselines, we further present numerical CHR results in Table~\ref{table: ce CHR}. Table~\ref{table: ce CHR} demonstrates that PPVF / SAGE / BESTFIT consistently attain the highest CHR compared to other caching baselines due to our efficient utility prediction algorithm.
PPVF is slightly better than SAGE and BESTFIT in terms of the CHR performance among the small capacities, while BESTFIT achieves the highest CHR when the caching capacity is large.

\end{appendices}